\documentclass{aa}

\usepackage[varg]{txfonts} 
\usepackage{graphicx} 
\usepackage{natbib}
\bibpunct{(}{)}{;}{a}{}{,} 
\usepackage{hyperref}

\newcommand*{\fullref}[1]{\hyperref[{#1}]{\autoref*{#1} \nameref*{#1}}}
\newcommand{\change}[1]{#1}

\defcitealias{Mishra2022a}{Paper I}
\def\paperone/{\citetalias{Mishra2022a}}
\graphicspath{{./images/}}

\def\kobe/{\texttt{KOBE}}
\def\kobeshadows/{\texttt{\kobe/-Shadows}}
\def\kobetransits/{\texttt{\kobe/-Transits}}
\def\kobevetter/{\texttt{\kobe/-Vetter}}
\def\pip/{peas in a pod}
\def\figwidth{8cm}
\def\similar/{similar}
\def\mixed/{mixed}
\def\antiordered/{anti-ordered}
\def\ordered/{ordered}

\def\mearth{M_\oplus}

\def\mjupiter{M_\text{J}}

\def\msun{M_\odot}

\def\sigmas0{\Sigma_\mathrm{s,0}}

\def\fpg{f_{\rm D/G}}
\def\fpgsun{f_\mathrm{D/G,\odot}}
\def\feh{[\mathrm{Fe/H}]}

\def\periodictce/{\text{\textit{p}TCE}}

\def\nplanet{n}

\def\cvtext/{coefficient of variation}
\def\cstext/{coefficient of similarity}
\def\cs{C_{S}}
\def\cv{C_{V}}

\hypersetup{pdfauthor={Lokesh Mishra}, 
	pdftitle=Framework for architecture of exoplanetary systems II,  
	colorlinks=true, 
	urlcolor=blue, 
	linkcolor=blue,  
	citecolor=blue}

\usepackage{orcidlink}

\begin{document}
	
	\title{A framework for the architecture of exoplanetary systems}
	\subtitle{II. Nature versus nurture: Emergent formation pathways of architecture classes}
	
	\author{Lokesh Mishra\inst{\ref{unibe},\ref{unige}}\orcidlink{0000-0002-1256-7261}
		\and Yann Alibert\inst{\ref{unibe}}\orcidlink{0000-0002-4644-8818}
		\and St\'ephane Udry\inst{\ref{unige}} \orcidlink{0000-0001-7576-6236}
		\and Christoph Mordasini\inst{\ref{unibe}}\orcidlink{0000-0002-1013-2811}
	}
	
	\authorrunning{L. Mishra et al.}
	\titlerunning{Architecture Framework II -- Nature versus nurture: Emergent formation pathways of architecture classes}
	
	\institute{
		Institute of Physics, University of Bern, Gesellschaftsstrasse 6, 3012 Bern, Switzerland\label{unibe}
		\\\email{\hyperref{mailto:exomishra@gmail.com}{}{}{exomishra@gmail.com}}
		\and
		Geneva Observatory, University of Geneva, Chemin Pegasi 51b, 1290 Versoix, Switzerland\label{unige}       
	}

	\date{Received DD MMM YYYY; accepted DD MMM YYYY}
	
	
	\abstract{In the first paper of this series, we proposed a model-independent framework for characterising the architecture of planetary systems at the system level. There are four classes of planetary system architecture: similar, mixed, anti-ordered, and ordered. In this paper, we investigate the formation pathways leading to these four architecture classes. To understand the role of nature versus nurture in sculpting the final (mass) architecture of a system, we apply our architecture framework to  synthetic planetary systems --- formed via core-accretion --- using the Bern model. General patterns emerge in the formation pathways of the four architecture classes. Almost all planetary systems emerging from protoplanetary disks whose initial solid mass was less than one Jupiter mass are similar. Systems emerging from heavier disks may become mixed, anti-ordered, or ordered. Increasing dynamical interactions (planet--planet, planet--disk) tends to shift a system's architecture from mixed to anti-ordered to ordered. Our model predicts the existence of a new metallicity--architecture correlation. Similar systems have very high occurrence around low-metallicity stars. The occurrence of the anti-ordered and ordered classes increases with increasing metallicity. The occurrence of mixed architecture first increases and then decreases with increasing metallicity. In our synthetic planetary systems, the role of nature is disentangled from the role of nurture. Nature (or initial conditions) pre-determines whether the architecture of a system becomes similar; otherwise nurture influences whether a system becomes mixed, anti-ordered, or ordered. We propose the `Aryabhata formation scenario' to explain some planetary systems which host only water-rich worlds. We finish this paper with a discussion of future observational and theoretical works that may support or refute the results of this paper.}

	

	\maketitle
	
	\section{Introduction}
	\label{sec:introduction}
	
	Studying planetary systems as single units of a physical system makes them amenable to system level examinations. Investigating the ensemble of bound objects (host star(s), planets, minor bodies) coherently can allow a deeper and more comprehensive understanding of exoplanetary astrophysics to emerge. The purview of this multi-body physics covers a breadth of topics including stability of planetary systems \citep{Gladman1993, Laskar1997, Laskar2000,Chambers1999, Fang2013, Pu2015,Laskar2017, Obertas2017,Petit2018, Wang2019, Yeh2020, Tamayo2020, Turrini2020}, stellar host and protoplanetary disk properties \citep{Petigura2018,Manara2019,Mulders2021}, novel approaches to system-level characterisation \citep{Tremaine2015,Kipping2018,Alibert2019,Mishra2019,Gilbert2020,Bashi2021, Sandford2021}, and the architecture of planetary systems \citep{Lissauer2011,Ciardi2013, Fabrycky2014,Weiss2018,Millholland2017, Adams2019,Adams2020,Mulders2020,He2019,He2020,Mishra2021,Adibekyan2021, Millholland2021, Winter2020}. 
	Analysing multi-body system level physics may allow us to understand whether planetary systems are self-organizing emergent structures -- i.e. whether global level patterns are emerging from local level interactions. 
	
	Inspired by the peas in a pod architecture \citep{Weiss2018, Millholland2017, Mishra2021}, we introduced a new framework for studying the architecture of planetary systems \citep[][; hereafter \paperone/]{Mishra2022a}. Studying the architecture as a global-system level phenomena, this framework allows us to characterise, quantify, and compare the architecture of individual planetary systems. Four classes of planetary system architecture emerged from this framework. These classes are labelled \similar/, \mixed/, \antiordered/, and \ordered/  depending on the arrangement and distribution of planets around the host star. The key idea behind this framework is that the arrangement and distribution of planets contains additional information that cannot be extracted by studying single planets individually. Hints of the presence of this additional information were revealed in some works \citep{Tremaine2015, Laskar2017,Kipping2018, Mishra2019, Gilbert2020, Sandford2021}. 
	
	Explaining the formation, evolution, and final assembly of planetary systems remains an outstanding theoretical problem. Planet-formation physics spans astronomical orders of magnitude in mass, size, and time \citep{Udry2007,2010apf..book.....A}. The processes occurring during planet formation convert gases and micron-sized dust particles from the protoplanetary disk into different kinds of planets arranged in different architectures over timescales of millions and billions of years. However, it remains unclear as to how initial conditions derived from the host star or protoplanetary disk combine with the formation and evolution processes to give rise to the observed exoplanetary systems. 
	
	We are interested in understanding the role of {nature} versus {nurture} in sculpting the final planetary system and the extent to which the character of the mature planetary system is influenced by its initial conditions. \cite{Kipping2018} suggested, using an entropy-like formulation for planetary systems, that the initial conditions of planet formation could be inferred based on their present-day architecture. However, the presence of stochastic processes makes it difficult to connect the initial conditions with the final system. It is also unclear as to whether or not stochastic physical processes can erase all memory of initial conditions, or indeed leave their own impressions on the final architecture. Using ideas from the fields of machine-learning-based natural language processing, \cite{Sandford2021} showed that planetary systems are not randomly assembled. While it is clear that planetary systems are not identical copies of one another, the quest for quantifying the similarity between planetary systems is a tantalising one. 
	
	In this paper, we investigate the formation pathways that lead to the four architecture classes. Due to the stochastic nature of this problem, understanding the formation of a single planetary system can be very complicated. For example, two systems with almost identical initial conditions may evolve into two completely different planetary systems. Chaos arising from multi-body gravitational interactions may cause differing formation pathways for these two systems. However, some patterns are found to emerge when studying planetary systems as part of an ensemble. These trends, as we show in this paper, help us to understand the role played by initial conditions and physical processes in shaping the architecture. 
	
	\begin{figure}
		\resizebox{\hsize}{!}{\includegraphics[trim=0.5cm 0.5cm 2cm 0.5cm , clip]{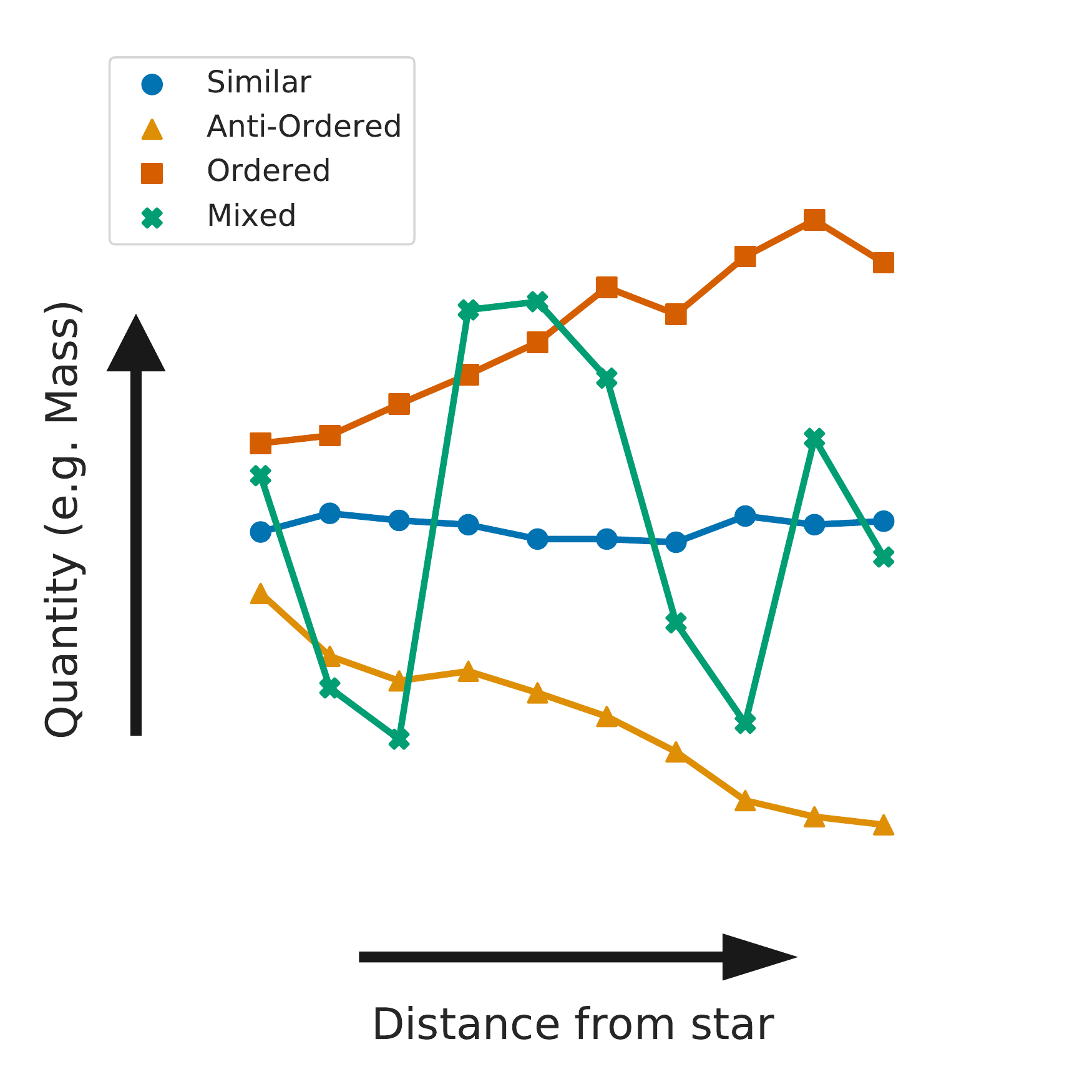}}
		
		\resizebox{\hsize}{!}{\includegraphics[trim=0.5cm 0cm 0.2cm 0cm,clip]{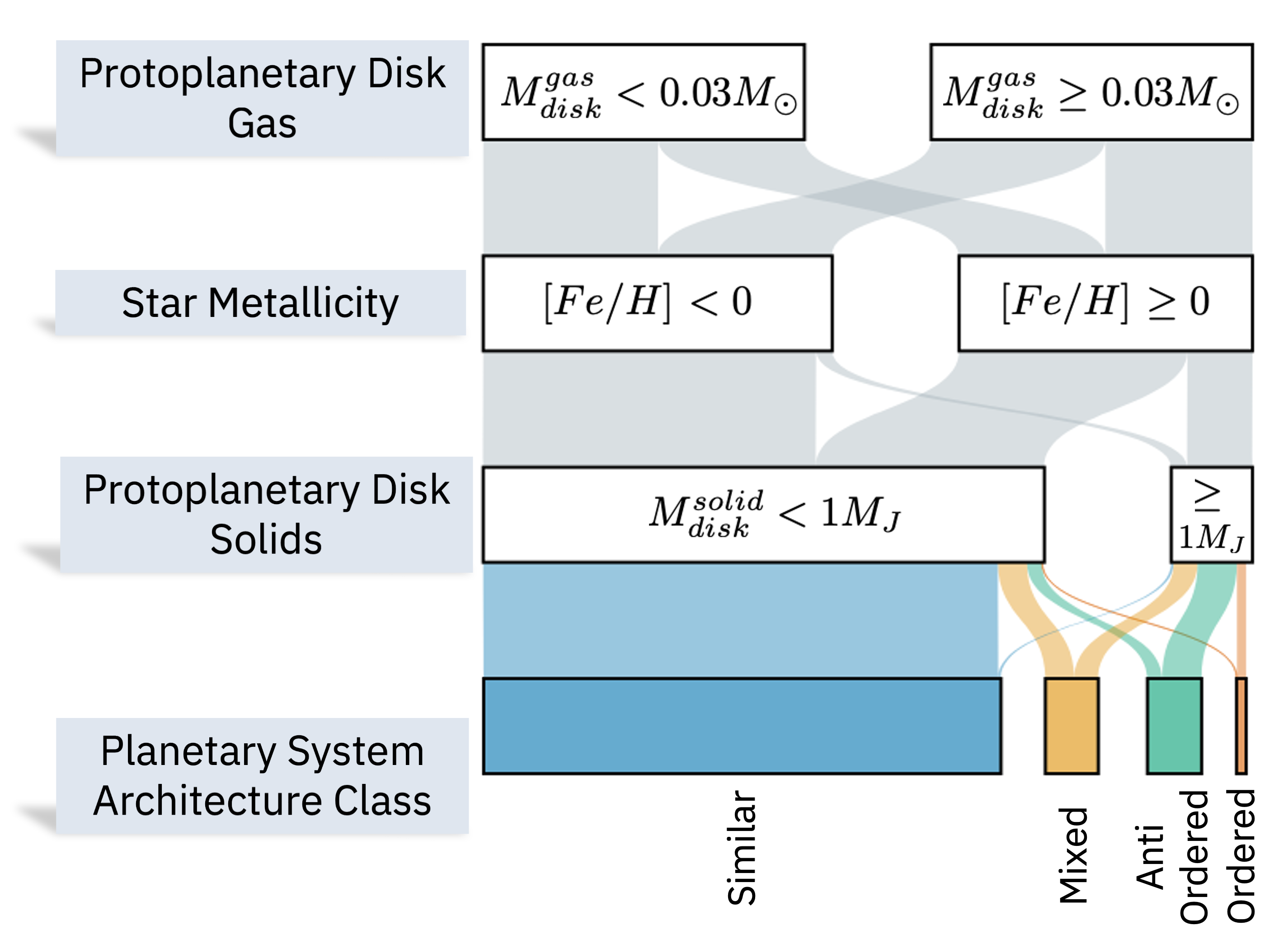}}
		\caption{The four classes of planetary system architecture and their emergent formation pathways. 
			\\
			\textit{Top:} Reproduced from \paperone/ -- Schematic diagram depicting the Four classes of planetary system architecture: \similar/, \antiordered/, \mixed/, and \ordered/. Depending on how a quantity (such as mass or size) varies from one planet to another, the architecture of a system can be identified. The framework is model independent. 
			\\
			\textit{Bottom:} Emergence of formation pathways: Sankey diagram depicting the emergence of formation pathways of architecture classes. The thickness of the links and nodes is proportional to the relative number of synthetic systems in our simulation. This result is derived from synthetic planetary systems around a solar mass star via the Bern model. Disk gas mass and metallicity are binned at their median values.}
		\label{fig:schematic_idea}
	\end{figure}
	
	\change{\change{Figure \ref{fig:schematic_idea} (bottom) summarises the main findings of this paper. We show that the effects of planet formation and evolution processes are imprinted in the system-level architecture. Figure 1 shows the formation pathways of the architecture classes that emerge due to the system-level approach of our architecture framework (Fig. \ref{fig:schematic_idea} (top)). This sankey diagram has nodes for protoplanetary disk gas mass, protoplanetary disk solid mass, metallicity, and planetary architecture.}}
	We find that the formation of \similar/  planetary systems is dominated by initial conditions. If the initial conditions disfavour the formation of \similar/ architecture, the other three architectures may emerge. Whether the final architecture is \mixed/, \ordered/, or \antiordered/ seems to depend on the stochastic formation processes. Increasing dynamical interactions (disk--planet, planet--planet) generally tends to produce \mixed/, \antiordered/, and then \ordered/ architectures, respectively.
	
	We first summarise the architecture framework and some results from \paperone/ in Sect. \ref{sec:summary}. We study the role of nature (initial conditions) and nurture (dynamical processes) in Sects. \ref{sec:nature} and \ref{sec:nurture}, respectively. In these sections, we study the influence of protoplanetary disk mass, metallicity, protoplanetary disk lifetime, planet--disk interactions, planet--planet interactions, and \textit{N}-body interactions on the final architecture of simulated planetary systems. We summarise our results, suggest possible future studies, and conclude this paper in Sect. \ref{sec:conclusions}.
	
	\section{Summary of \paperone/ and the Bern model}
	\label{sec:summary}
	
	\subsection{Architecture framework}
	
	The arrangement of multiple planets and the collective distribution of their physical properties around the host star(s) characterises the architecture of a planetary system \citep{Mishra2021}. To quantify the architecture of a planetary system, we developed a novel model-independent framework \paperone/. \change{Some key aspects of this framework are briefly summarised here, and we refer the reader to Sect. 3 of \paperone/ for details.}
	
	Conceptually, the framework defines four classes of planetary system architecture: \similar/, \mixed/, \antiordered/, and \ordered/. Consider a planetary quantity (such as mass, radius, etc.) as a function of the distance of the planet to the host star \change{(see Fig. \ref{fig:schematic_idea})}. When all planets in a system have similar values of a planetary quantity, the architecture of such systems is similar. When the planetary quantity increases with increasing distance, the system is said to exhibit an \ordered/ architecture. Alternatively, if the quantity shows an overall decreasing trend with increasing distance, the architecture is considered to be \antiordered/. Finally, the planetary quantities could also show variations that are not captured in the three classes above. A \mixed/ architecture may depict large, increasing, and decreasing variations with distance. By studying the variation of a planetary quantity with distance for all planets in the system, our framework captures the arrangement and distribution of planets in the system.
	
	The architecture of a system is quantified via two coefficients: the \cstext/, $\cs (q_i)$, and the \cvtext/, $\cv (q_i)$. Here, $q_i$ represents a planetary quantity (e.g. mass, radius, eccentricity, density) for the $i^\text{th}$ planet. When the coefficients are calculated using planetary masses, they inform us about the mass architecture of a system, that is, the arrangement and distribution of mass in a given system. Likewise, we can study the radius architecture, density architecture, water-mass-fraction architecture, eccentricity architecture, and so on. The versatility of our architecture framework lies in its ability to allow us to study the multifaceted architectures of a planetary system. In \paperone/, we explored the relationship between these different kinds of architectures. As in \paperone/, we identify the architecture of a system by its \change{bulk mass} architecture. 
	
	Calibrated on planetary masses, a classification scheme to identify the architecture class was proposed in \paperone/ (eq. 8). The $\cs$ versus $\cv$ plane represents the architecture space for planetary systems \change{(Fig. 3 in \paperone/)}. This new parameter space was found to be endowed with a curious mathematical property, namely planetary systems cannot occupy all parts of the architecture plane, as some regions of this parameter space are mathematically forbidden. 
	
	To understand the implications of this architecture framework, we applied it on several catalogues in \paperone/. These included 41 observed multi-planetary systems and numerically simulated systems via population synthesis using the Generation III Bern model \citep{Emsenhuber2021A, Emsenhuber2021B}.
	
	\subsection{Bern model}
	
	For the synthetic planetary systems, as the initial conditions and the physical processes are known, it is possible (and desirable) to understand how different architecture classes are formed. As this paper is dedicated to planet formation and its imprints on architecture, we briefly review the ingredients of the Bern model here. Readers interested in further details of this model are referred to the recent NGPPS series of papers \citep{Emsenhuber2021A,Emsenhuber2021B,Schlecker2021a,Burn2021,Schlecker2021b, Mishra2021}. The historic development of the Bern model may be traced through the works of \cite{Alibert2004, Alibert2005, Mordasini2009, 2011A&A...526A..63A, Mordasini2012(MR),Mordasini2012(models), Alibert2013, Fortier2013, Marboeuf2014a, Thiabaud2014, Dittkrist2014, 2014ApJ...795...65J} and is reviewed in \cite{Benz2014, Mordasini2018}.
	
	\change{Based on the core-accretion paradigm \citep{Pollack1996},} the Bern model is a global model of planet formation and evolution. The model studies the growth of several lunar-mass protoplanetary embryos embedded in protoplanetary disks (consisting of a gaseous and solid phase) around a solar-type star. \change{The disk model is based on viscous angular momentum transport \citep{Lynden-Bell1974, Veras2004, Hueso2005}. Turbulence is characterised by the \citet{Shakura1973} approach. The initial mass of the solid disk depends on the metallicity of the star and also on the condensation state of the molecules in the disk \citep{Thiabaud2014}. The solids in the disk are composed of a swarm of rocky and icy planetesimals. The solids in the disk evolve via (a) accretion by growing planets, (b) interaction with gaseous disk, (c) dynamical stirring from planets and other planetesimals, and so on \citep{Fortier2013}. The 1D geometrically thin disk evolution is studied up to 1000 au.} 
	
	This star--disk--embryo numerical system is endowed with several physical processes, which are occurring simultaneously and in a self-consistently coupled way. Some of these physical processes are: stellar evolution \citep{Baraffe2015}, interactions between viscous protoplanetary disk and star \citep{Lynden-Bell1974, Shakura1973, Clarke2001, Matsuyama2003, Veras2004, Nakamoto1994,  Hueso2005}, condensation of volatile and/or refractory species \citep{Marboeuf2014a, Marboeuf2014b, Thiabaud2014}, planet formation physics \citep{Alibert2013, Fortier2013, Mordasini2012(models)}, orbital and tidal migration \citep{Coleman2014, Paardekooper2011, Dittkrist2014}, gravitational \textit{N}-body interactions \citep{Chambers1999, Alibert2013, Emsenhuber2021A, Emsenhuber2021B}, atmospheric escape \citep{2014ApJ...795...65J}, bloating \citep{2021A&A...645A..79S}, and so on (see Fig. 1 in \cite{Mishra2019} for a schematic diagram). In addition, the model also calculates the internal structure of all planets, assuming them all to be spherically symmetric. 
	
	In the synthetic planetary population we use in the present work, some initial conditions are fixed, namely we use a $1 \msun$ mass star and a disk viscosity $\alpha = 2 \times 10^{-3}$, describing the initial shape of the gas and planetesimal disks via power laws \citep{Veras2004}, with a planetesimal size of $300 m$, and a fixed density (rocky 3.2 $\text{g cm}^{-3}$, icy 1 $\text{g cm}^{-3}$). \change{We add 100 protoplanetary embryos to the protoplanetary disk. We ensure that no two embryos start within 10 hill radii of each other \citep{Kokubo1998, Kokubo2002}.} This model is then run 1000 times while varying other initial conditions. We varied the initial gas mass in the protoplanetary disk, disk lifetime, stellar metallicity, disk inner edge, and the initial location of the protoplanetary embryos \citep[for details see][]{Emsenhuber2021B}. 
	
	The Bern model includes a significant variety of physics and uses plausible choices of initial conditions, which are motivated by observations. However, it is only a simplified low-dimensional  approximation of our current understanding of planet formation. For example, we model planet formation via core-accretion only and ignore other methods, such as disk instability \citep{Schib2021}. Among others, we also assume that the dust-to-gas ratio is the same for both the host star and the disk, and that all dust in the disk is aggregated into planetesimals. The \textit{N}-body interactions are tracked for only 20 Myr, which may be inadequate to capture dynamical effects occurring in the outer parts of the system. The assumptions, choices, and simplifications made in this model may have a strong impact on the outcome of this paper. Nevertheless, exploring the implications of our architecture framework using synthetic populations via the Bern model is a necessary first step. The main result of this paper is not in understanding the formation of any single planetary system but to show that, for different architecture classes, \change{discernible patterns of formation pathways emerge}. Future studies could apply our architecture framework (from \paperone/) with other planet formation models. If the formation pathways for the different architecture classes were found to remain the same after using different formation models, then our results would be strengthened and become more robust.
	
	\section{{Nature:} Role of star and disk initial conditions}
	\label{sec:nature}

	In this section, we study the connection between the initial conditions and the final architecture of a system. We begin by counting the number of different architecture classes that emerge from our population synthesis as a function of the various initial conditions that are varied. The role of varying disk masses and stellar metallicities is presented in Sect. \ref{subsec:mass_metallicity}, and that of varying disk lifetimes in Sect. \ref{subsec:disklifetime}. For completeness, we measure the relative count for an architecture class within a bin by dividing the number of systems of a particular architecture class in a bin by the total number of systems in that bin. \change{We emphasise that, as in \paperone/, the architecture of a system is identified with its bulk mass architecture. Thus, when we refer to a \similar/ or \ordered/ system, we are referring to a system whose bulk mass architecture is \similar/ or \ordered/, respectively.}
	
	\subsection{Protoplanetary disk: Mass and stellar metallicity}
	\label{subsec:mass_metallicity}
	
	\def\figwidtha{7cm}
	\begin{figure*}
		\centering
		\includegraphics[width=\figwidtha]{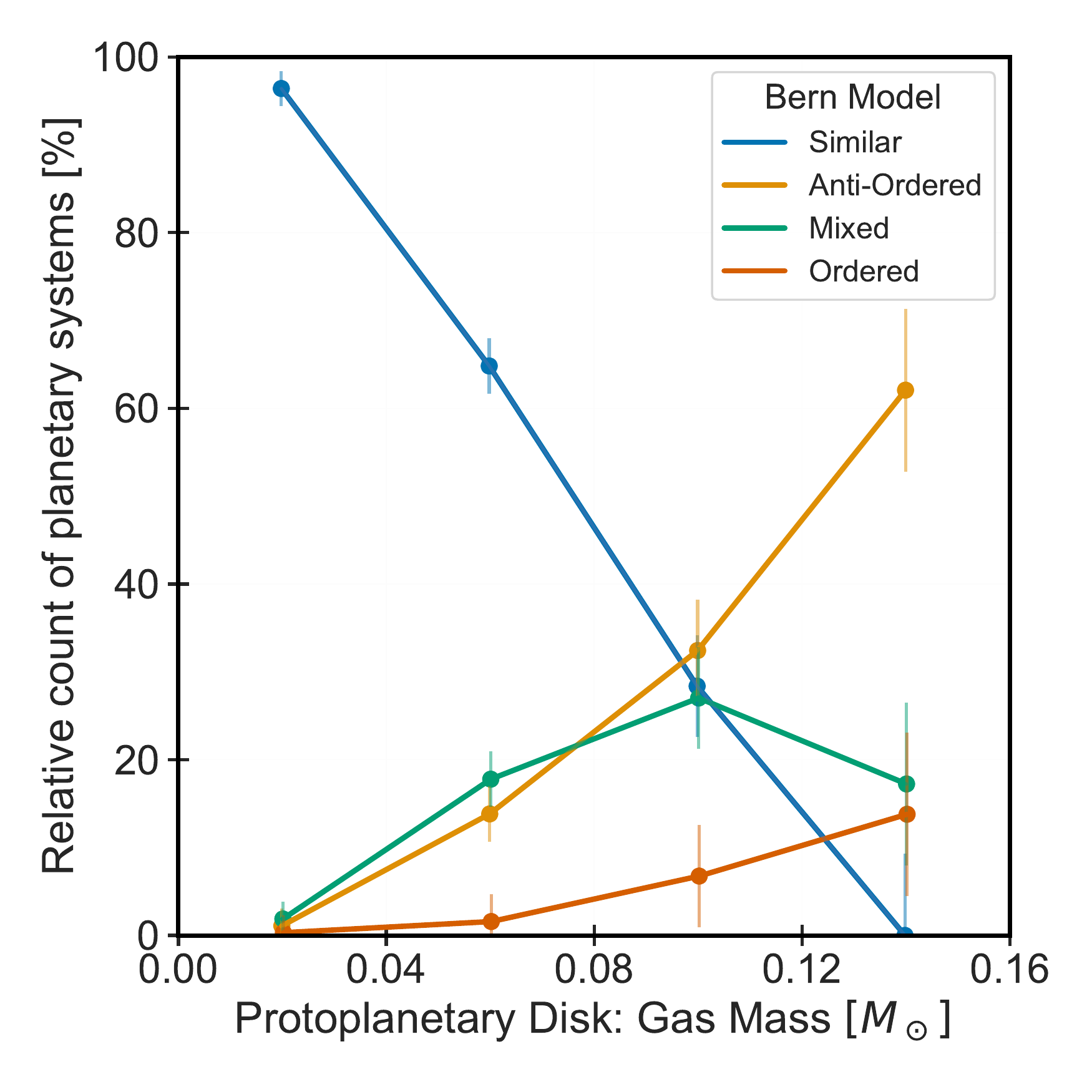}          
		\includegraphics[width=\figwidtha]{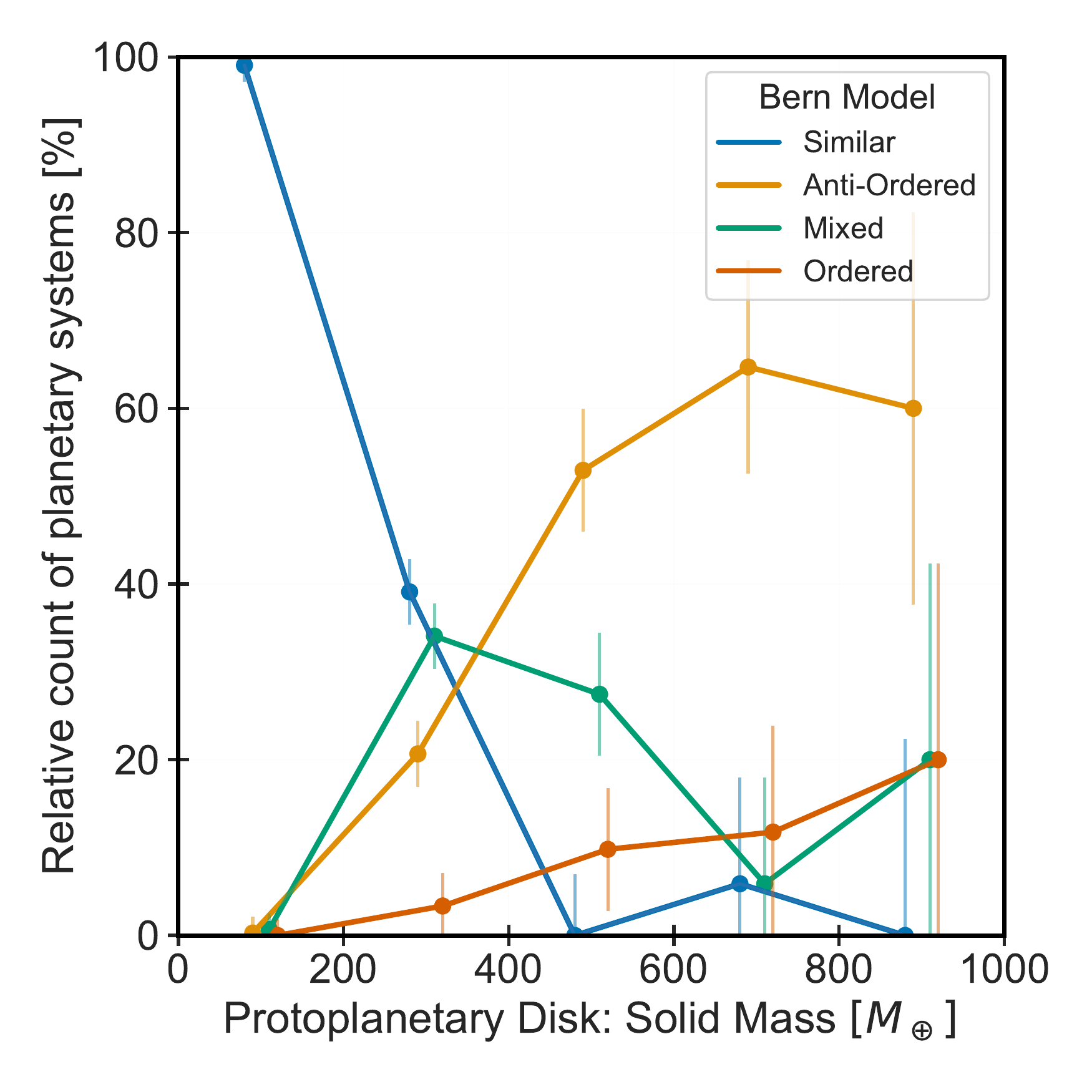}
		\includegraphics[width=\figwidtha]{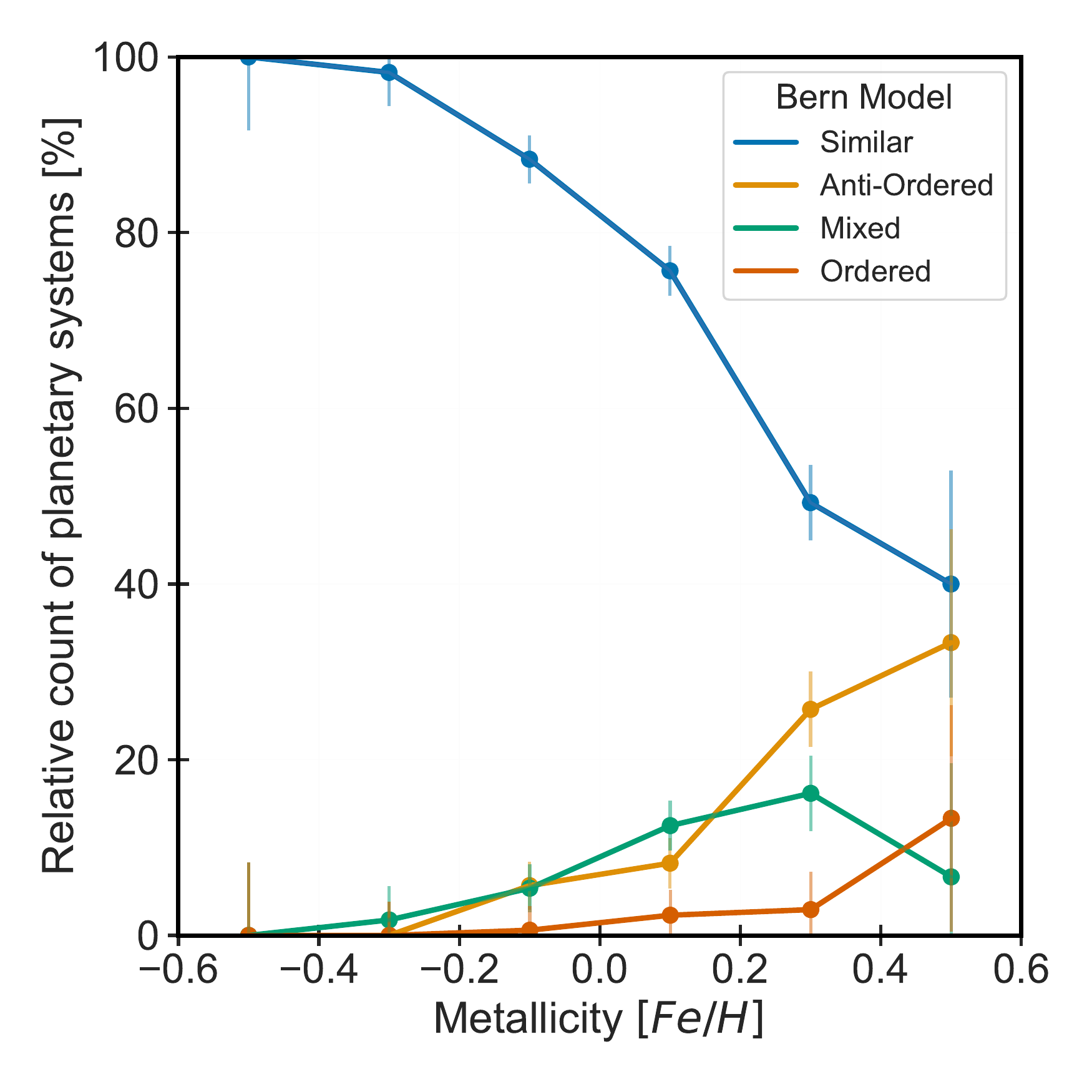}                       \includegraphics[width=\figwidtha]{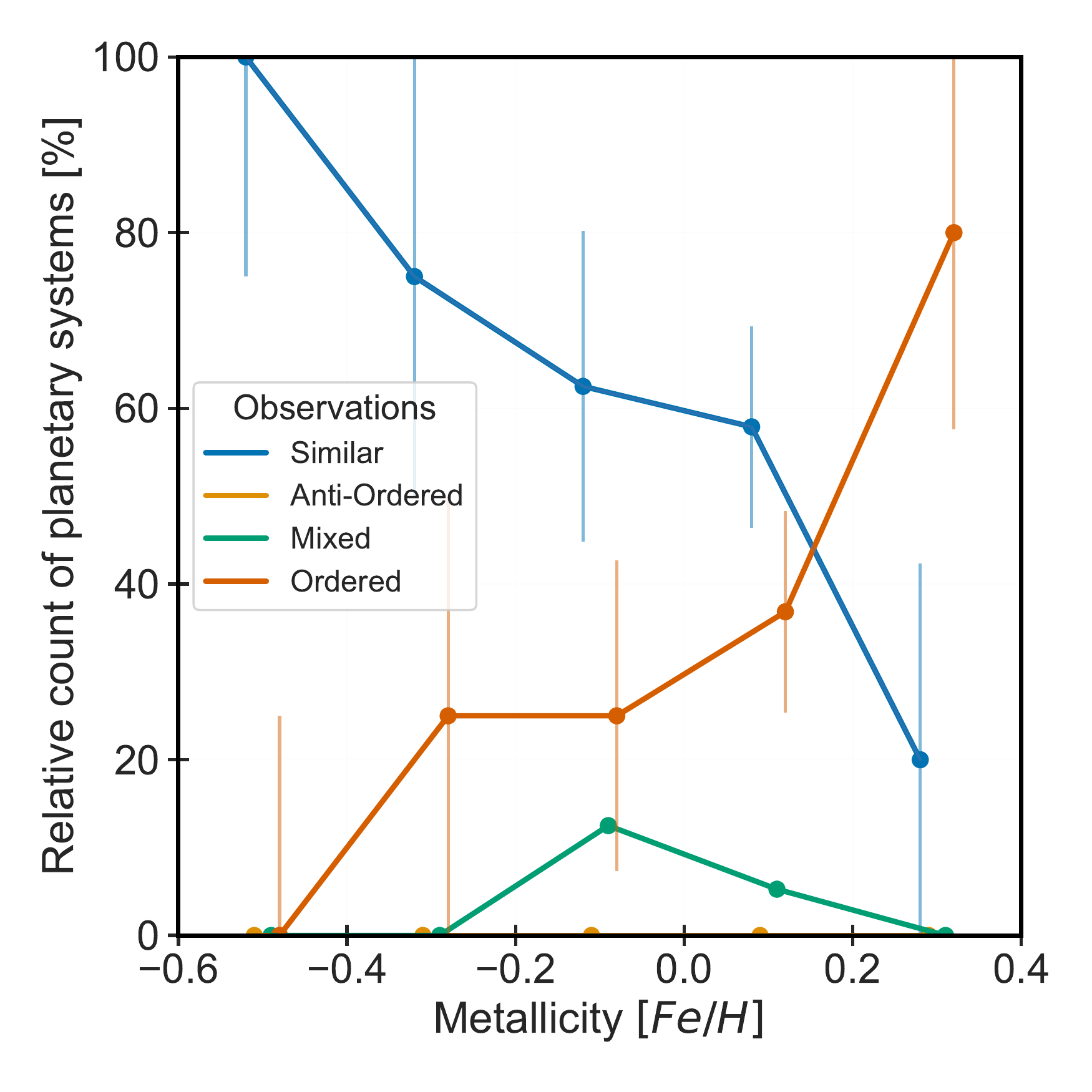}
		\includegraphics[width=\figwidtha]{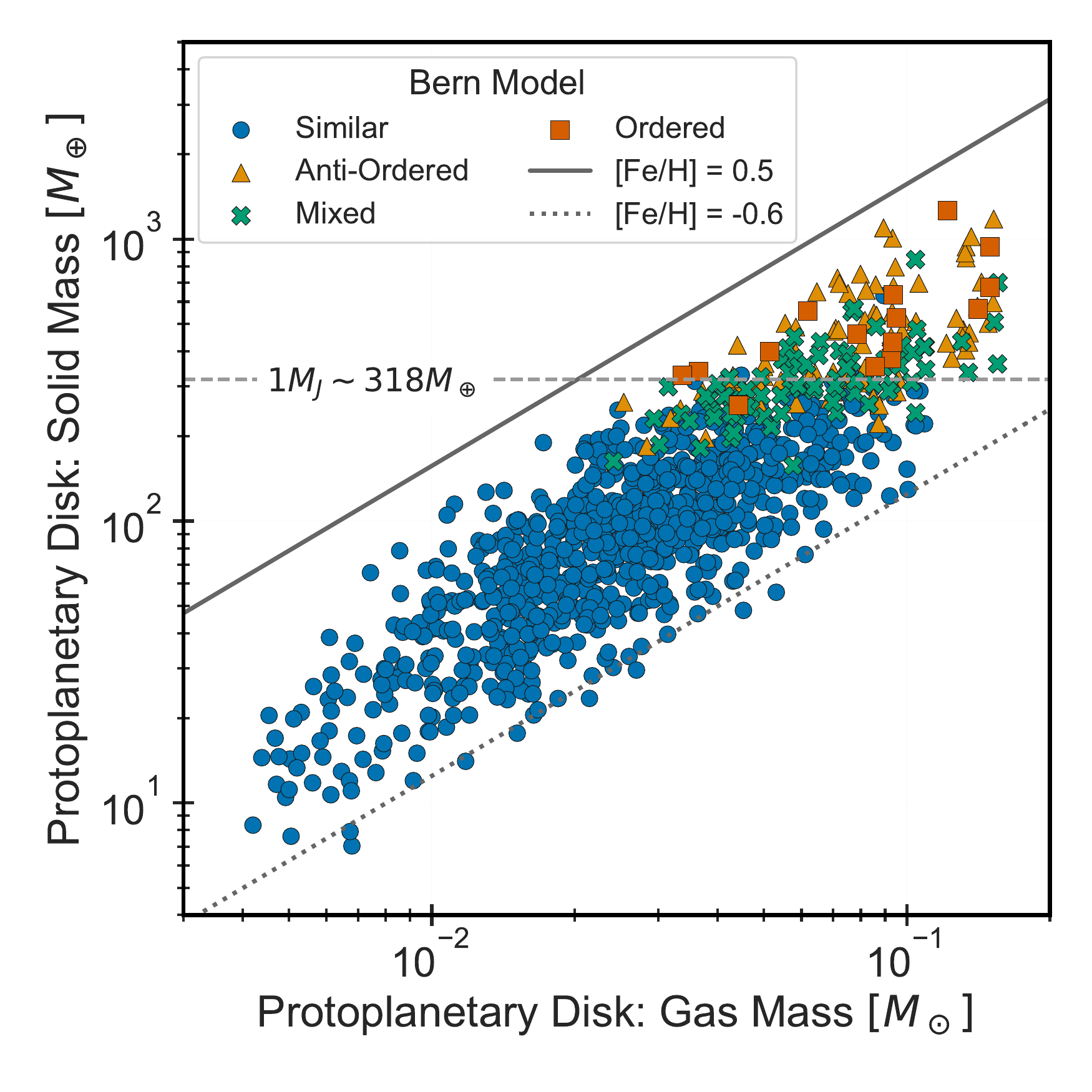}              
		\includegraphics[width=\figwidtha]{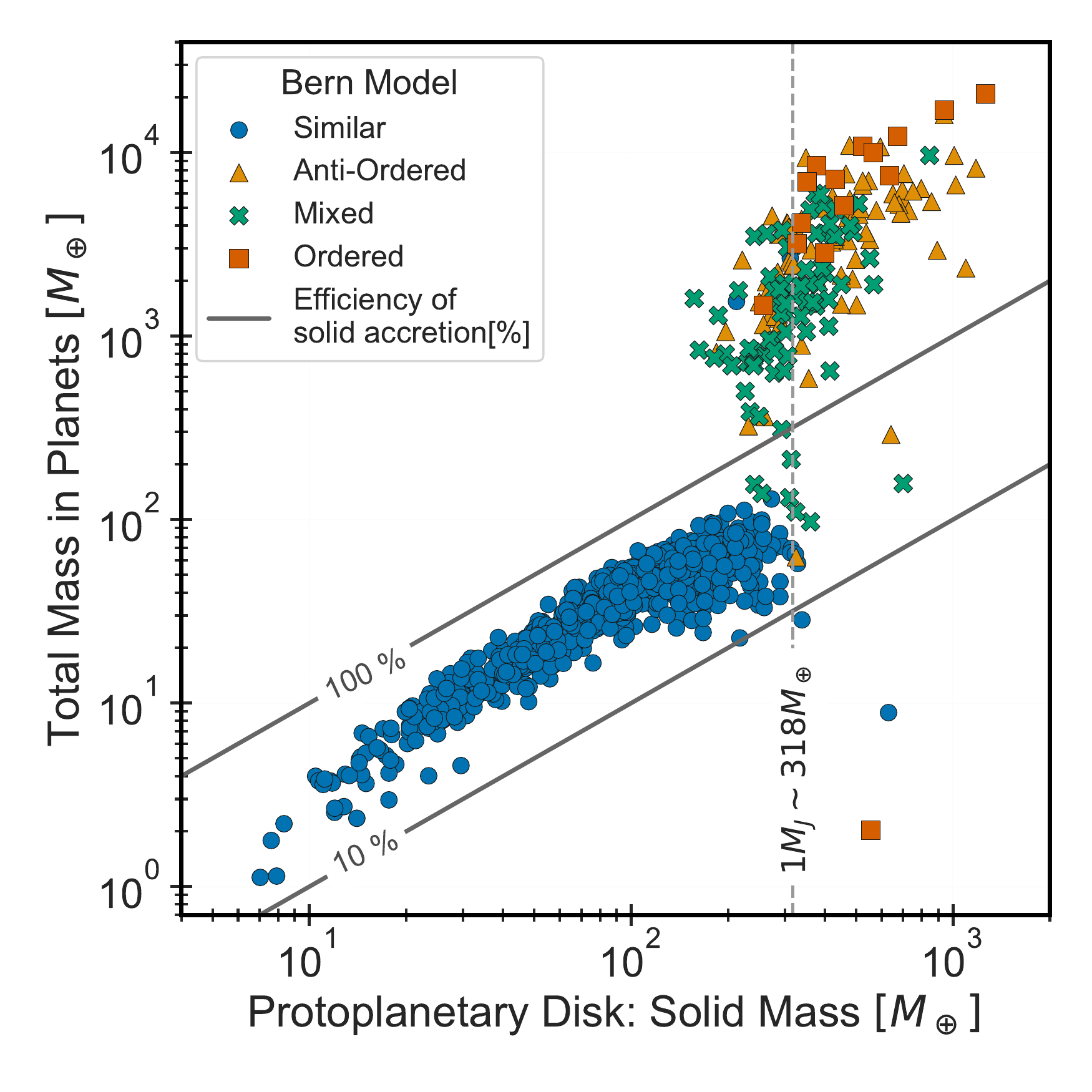}
		\caption{Role of disk mass and the metallicity--architecture correlation. The top two rows show the binned relative count of each architecture class as a function of initial disk gas mass (upper left), disk solid mass (upper right), stellar metallicity in the synthetic population (middle left), and stellar metallicity in observed systems (middle right). The length of the error bars corresponds to the total number of systems in each bin as: ${100}/{\sqrt{\mathrm{bin\ counts}}}$.
			In the bottom panels, each point corresponds to a single planetary system. The system architecture is indicated by the colour and shape of the marker. 
			The bottom left panel shows the solid mass in the disk as a function of the disk gas mass. The two diagonal lines convey the role of stellar metallicity. The dashed horizontal line indicates the mass of Jupiter. 
			The bottom right panel shows the total mass in planets as a function of the solid mass in the protoplanetary disk. The two diagonal lines indicate the efficiency of converting solids from the disk into planets. If the planets in a hypothetical system could accrete all the solid mass of its disk, and these planets had no gaseous atmosphere, then  such a system would lie on the diagonal line corresponding to $100\%$ accretion efficiency. The dashed vertical line indicates the mass of Jupiter.
		}
		\label{fig:bincounts_mass}
	\end{figure*}
	
	Figure \ref{fig:bincounts_mass} (upper left) shows the dependence of the architecture class relative counts on the initial mass of gas in the protoplanetary disk. Over $96\%$ of all disks that started with gas masses $\lesssim 0.04 \msun$ give rise to planetary systems of \similar/ architecture. About $1\%$ of these low-mass disks lead to each of the other three architecture classes. The relative count of systems with \similar/ architecture shows a clear decreasing trend with increasing mass in the disk gas. 
	
	The production of the remaining three architecture classes tends to increase with increasing disk gas mass, but with distinct trends. As the mass in the gas disk increases, the relative count of \mixed/ architectures increases first, and then decreases for gas mass $\gtrsim 0.12 \msun$. The relative count for both \antiordered/ and \ordered/ architectures continues to increase with increasing disk mass. Anti-ordered architectures become the most common outcome from large disks with gas mass $\gtrsim 0.12 \msun$.
	
	In Fig. \ref{fig:bincounts_mass} (upper right), we see the binned relative count of different architecture classes as a function of the mass of the solids in the protoplanetary disk. This plot shows some of the same features that we saw in Fig. \ref{fig:bincounts_mass} (upper left). About $99\%$ of all disks that have solid masses $\lesssim 200 \mearth$ give rise to \similar/ planetary systems. The production of \similar/ architecture decreases as the mass of solids in a disk is increased. 
	
	Before continuing, we note that this is already a result of considerable importance. The physical processes encoded in the Bern model are the same for all 1000 planetary systems. The only difference between these synthetic systems arises from the variations in their initial conditions. We are seeing that almost all low-mass disks give rise to only one architecture, the \similar/ class. This occurs despite all the physical processes that can act upon the system and induce some architectural variation. As we show below, the low mass of the disk limits some of the physical processes that sculpt a system's architecture. We conclude that the production of systems of the \similar/ architecture class is dominated by initial conditions. 
	
	Close to $60\%$ of all observed systems in our multi-planetary systems catalogue (from \paperone/) are \similar/ in their mass architecture (\paperone/). For some of these \similar/ class systems (like Trappist-1, TOI-178, etc), if their formation is via core-accretion, our work may suggest strong limits on the initial mass of their protoplanetary disks.  
	
	The relative count of the other three architecture classes increases as the solid mass in the disk increases. The production of \mixed/ architectures peaks around disks of $\approx 1 \mjupiter$ and then decreases. The prevalence of \antiordered/ and \ordered/ architectures continues to increase with increasing disk mass. For heavy massive disks, \antiordered/ architecture is the most common outcome.
	
	Figure \ref{fig:bincounts_mass} (middle left) shows the relative count of each architecture class in the synthetic population as a function of stellar metallicity. Figure \ref{fig:bincounts_mass} (middle right) shows the same for the 41 observed multi-planetary systems. The selection criterion for our observed catalogue is detailed in \paperone/. We find an interesting correlation between the metallicity and the architecture of a system, hereafter referred to as the metallicity--architecture correlation, and note  the following trends. Over $98\%$ of all systems with Fe/H $< - 0.2$ are of \similar/ type. The relative count of \similar/ architecture decreases as the metallicity is increased. The relative counts of the other three architecture classes are below $5\%$ for metallicities $\leq - 0.2$. At different rates, the relative counts of \mixed/, \ordered/, and \antiordered/ classes increase with increasing metallicity. Our catalogue of observed planetary systems shows an encouragingly similar trend.
	
	
	Our observations catalogue suffers from detection biases and incompleteness. One way in which these limitations manifest is that we do not find any observed example of \antiordered/ architecture. The qualitative trend for the relative count of observed system architectures as a function of their stellar metallicity agrees with our synthetic systems. For example, the relative count of \similar/ observed systems  decreases with increasing metallicity. The relative count of \ordered/ architectures continues to increase with increasing metallicity. 
	
	To understand the origin of these correlations, we study the relation between initial disk mass (both in solids and gases), stellar metallicity, and the final architecture of the systems in our model. In the Bern model, the initial solid mass of the disk is a fraction of the initial gas mass of the disk. This fraction is correlated with the dust-to-gas ratio, which also depends on the gas mass itself because the location of different icelines depend on it. By simulating systems with varying dust-to-gas ratio ($\fpg$), we simulate systems around stars with different metallicities. This is due to the following relation:
	\begin{equation}
		\label{eq:fpg}
		10^{\feh} = \frac{\fpg}{\fpgsun}, \quad \fpgsun = 0.0149 \ \text{\citep{Lodders2003}}.
	\end{equation} 
	The metallicities in our simulations vary from $- 0.6$ to $0.5$ following \cite{Santos2005}.
	
	Figure \ref{fig:bincounts_mass} shows the solid disk mass as a function of the gas disk mass (bottom left) and the total mass in the planets as a function of the solid disk mass (bottom right). Each point represents one planetary system, and the shape and colour of the marker shows its final architecture. These two plots help us understand the correlations discussed above. 
	
	The bottom left panel of Fig. 2 shows the relationship between gas disk mass, solid disk mass, metallicity, and the final architecture of the system. Generally, when the mass of the solids in a disk is $\gtrsim 1 \mjupiter (\approx 318 \mearth)$, the production of architectures other than \similar/ is triggered. We note that up to a certain gas disk mass ($\lesssim 0.02 \msun$), irrespective of the metallicity, all disks lead to \similar/ architecture. For heavier gas disks ($\gtrsim 0.02 \msun$), metallicities begin to play a role. If the gas disk mass is high enough, even low metallicities ($\approx -0.2$) can trigger the production of architectures other than the \similar/ class. However, for lower gas disk masses, higher metallicities are required to produce about a $1 \mjupiter$ mass in the solid disk.
	
	It is clear that the mass in the solids of the protoplanetary disk plays an essential role here. The bottom right panel of Fig. \ref{fig:bincounts_mass} explains the above statement. The total mass in the planets increases as the mass of solids in the disk increases. When the mass of solids in the disk is $\sim 1 \mjupiter$, the distribution of total mass in planets shows a jump. This is because massive planets can begin to accrete significant amounts of gas. For the core-accretion scenario, this plot suggests that similar architectures occur for low-mass disks because they cannot produce massive giant planets. Gas giants are very effective in inducing dynamical stirring, which are in turn responsible for shaping the system architecture. This signifies the role played by physical processes in producing the \mixed/, \antiordered/, and \ordered/ architectures\footnote{The architecture framework is not sensitive to the absolute value of a planetary quantity, such as mass, but only the ratio of the quantities for adjacent planets. Independent of the architecture framework, we will present another system-level framework analysing the state of a planetary system. This other classification framework is sensitive to the absolute mass of a planet and will address the role of giant planets on system-level properties. The state classification framework reveals a drastic difference between systems with and without giant planets (Mishra et al. in prep.).}. 
	
	\subsection{Lifetime of the protoplanetary disk}
	\label{subsec:disklifetime}
	
	\begin{figure*}
		\centering
		\includegraphics[width=\figwidth]{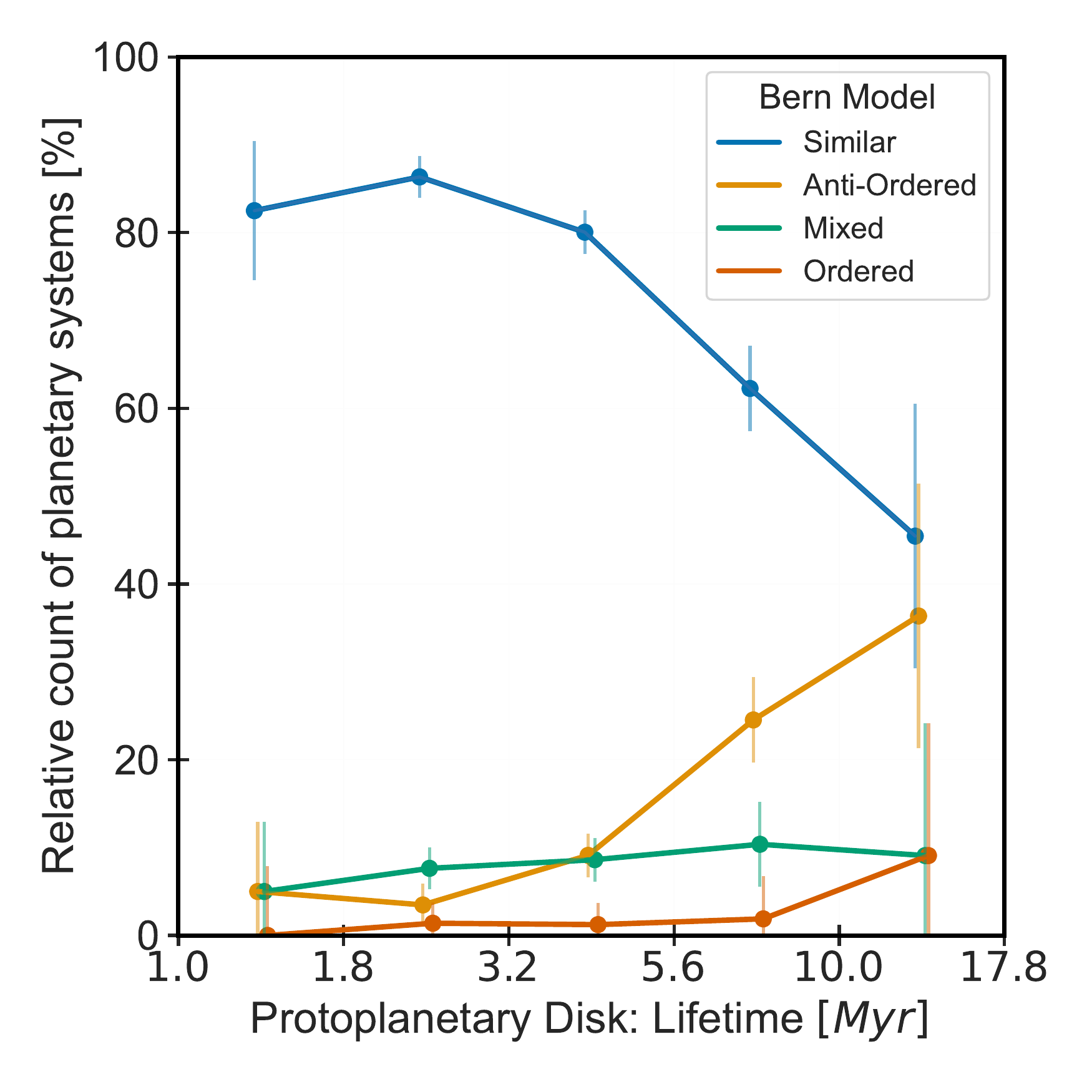}
		\includegraphics[width=\figwidth]{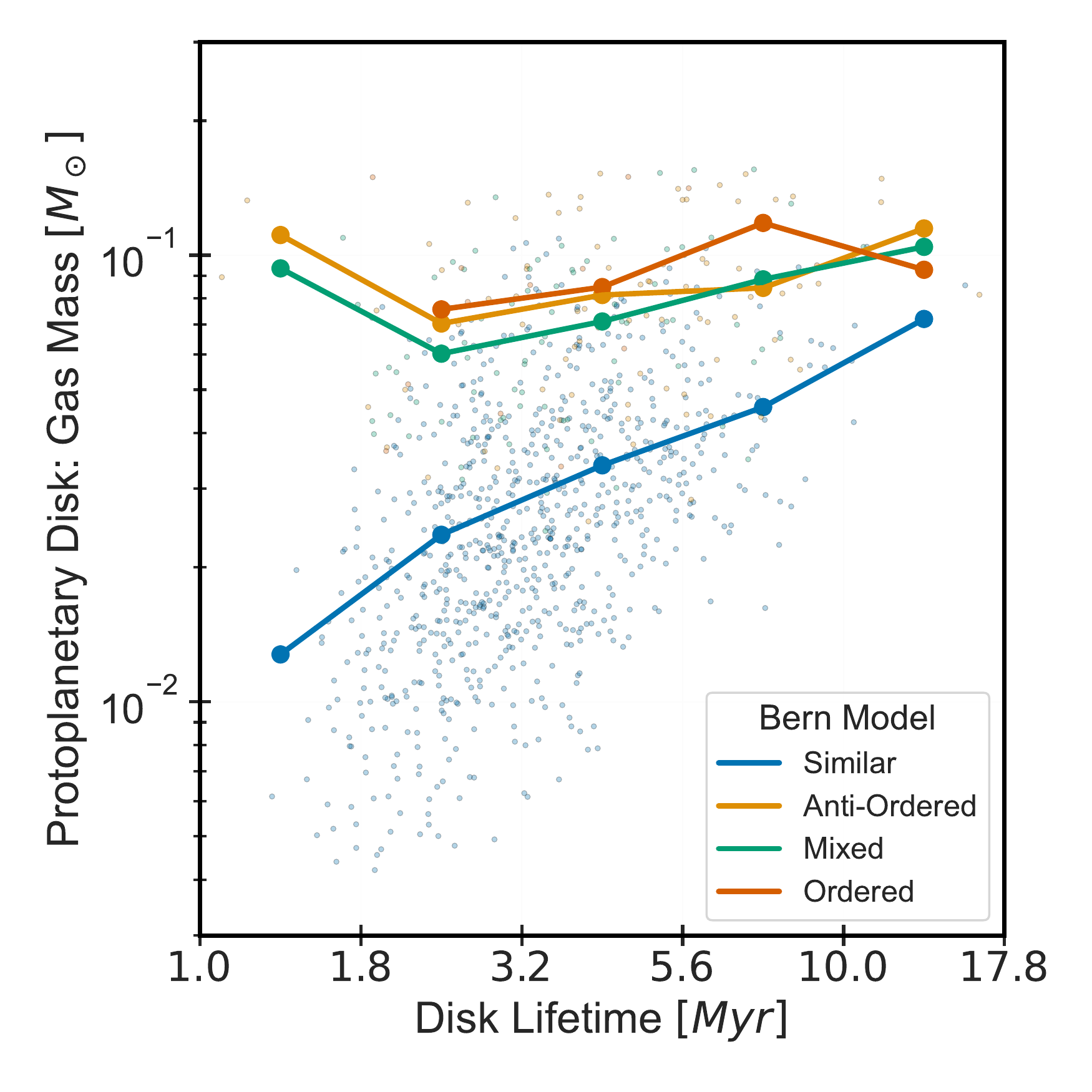}
		\caption{Role of disk lifetime on system architecture. Left: Binned relative counts of architecture classes as a function of disk lifetime. The length of error bars corresponds to the total number of systems in each bin, as: ${100}/{\sqrt{\mathrm{bin\ counts}}}$. Right: Scatter plot shows the disk gas mass as a function of disk lifetime. The solid lines show the binned average gas disk mass for each architecture class. }
		\label{fig:disklifetimecounts}
	\end{figure*}
	
	In this section, we explore the role of disk lifetime \change{(i.e. the age of a protoplanetary disk)} in defining the final architecture class of a system. The lifetime of a disk, in the Bern model, is influenced by the external disk photo-evaporation rate (see \cite{Emsenhuber2021A} for details) and the mass of the disk. 
	
	Figure \ref{fig:disklifetimecounts} (left) shows the binned relative count of system architecture as a function of disk lifetime.
	About $80\%$ of all disks with lifetimes ranging from 1 to 5 Myr produce systems of the \similar/ architecture class. The relative count of \similar/ systems decreases as disk lifetime increases. The relative count of \mixed/ architecture does not show any significant variation with disk lifetime. The relative counts of \antiordered/ and \ordered/ architectures vary as the disk lifetime increases. This suggests that the physical mechanisms by which disks shape the final architectures of systems play a role in shaping \similar/, \antiordered/, and \ordered/ architectures.
	
	The trends of the relative counts of architecture classes with disk lifetime are similar to the distribution of relative counts as functions of disk mass. We would like to understand whether system architecture is influenced by disk lifetime directly or via an inherent dependence of disk lifetime on disk mass. The right panel of Fig. \ref{fig:disklifetimecounts} shows the gas disk mass as a function of disk lifetime. The scatter plot depicting each individual disk shows that, generally, low-mass disks have short lifetimes. The solid lines depict the average gas mass for each architecture class for each disk lifetime bin. 
	
	The gas mass of the disks that go on to form systems of \mixed/, \antiordered/, or \ordered/ architecture shows  a weak dependence on disk lifetime. On average, the more massive disks seem to last longer. For disks that give rise to the \similar/ architecture class, this trend is clearly visible. If more massive disks also live longer, this partly explains the relative count distribution seen in Fig. \ref{fig:disklifetimecounts} (left). 
	
	However, disks also affect the planetary architecture in other interesting ways, namely orbital migration and eccentricity, and inclination damping. We study the effect of these planet--disk interactions in shaping system architecture in Sect. \ref{subsec:diskplanet}.
	
	
	\section{{Nurture:} Role of dynamical stirring}
	\label{sec:nurture}
	
	Whether or not the final architecture of a planetary system is pre-determined by its initial conditions from the host star and the protoplanetary disk remains unclear. If not, the mechanism by which dynamical processes shape the architecture of a planetary system remains to be determined. It also remains unclear as to whether or not dynamical processes remove all traces of initial conditions from the final system,  or whether these stochastic processes leave their impressions on the final architecture. In this section, we try to answer these questions. We focus our attention on dynamical interactions between planets and the protoplanetary disk, and the gravitational multi-body interactions amongst planets themselves. 
	
	While there exist several dynamical mechanisms that shape the final architecture, we simplify the task before us by concentrating on violent dynamical instabilities that change a planetary system in a non-trivial manner. For each synthetic planetary system, we count the number of planet--planet mergers,  planetary ejections, and planets falling into their host star. We use these counts as a proxy to assess the strength of dynamical interactions that occur in a system. In the subsequent subsections, we study planet--disk interactions and planet--planet interactions (mergers, ejections, stellar accretion). These dynamical effects give rise to stochasticity and are thereby inherently unpredictable. However, we hope that the underlying dynamical processes that are sculpting the system architecture emerge as patterns in the counts of these violent events.  
	
	\subsection{Planet--disk interactions}
	\label{subsec:diskplanet}
	\begin{figure*}[!h]
		\centering
		\includegraphics[width=\figwidth]{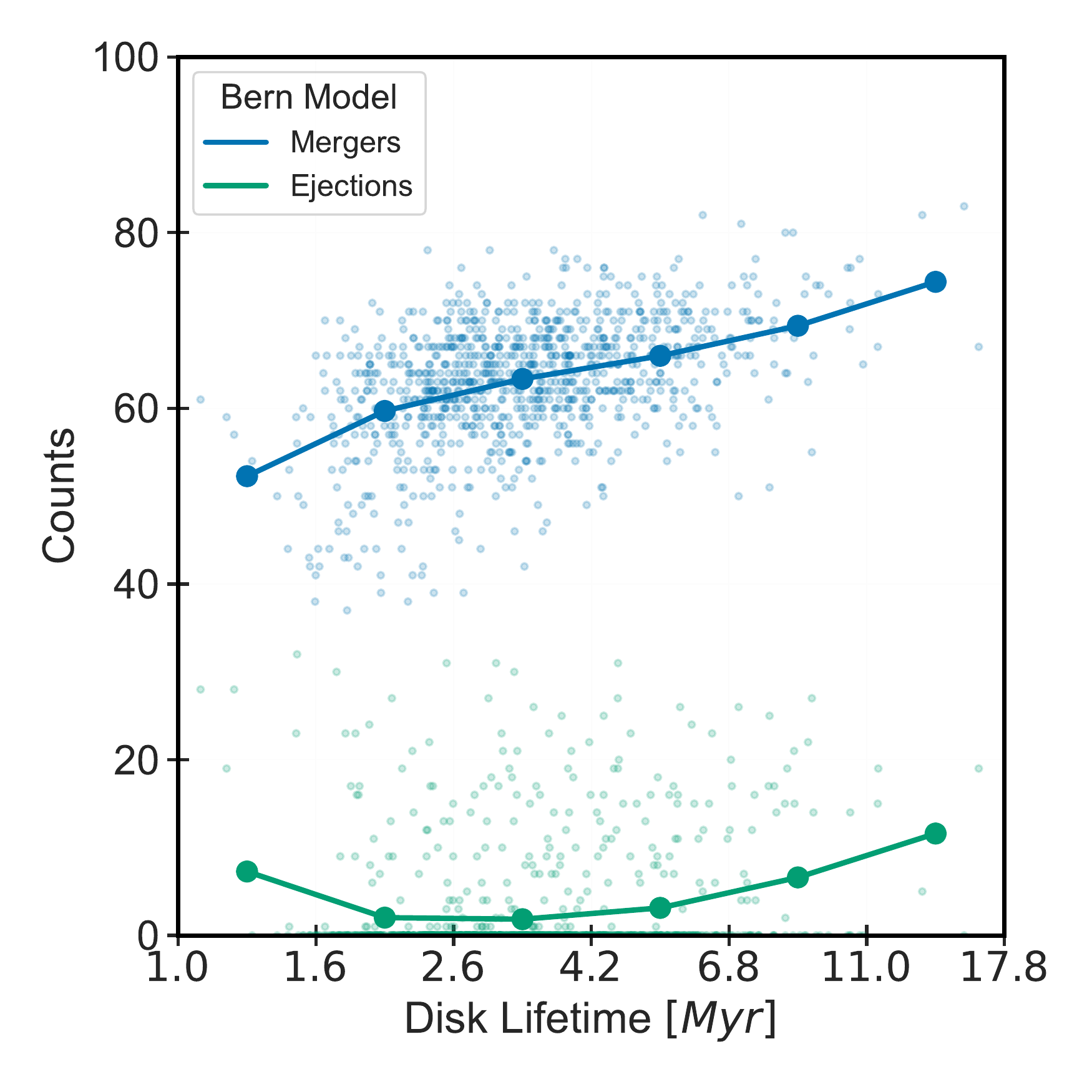}             
		\includegraphics[width=\figwidth]{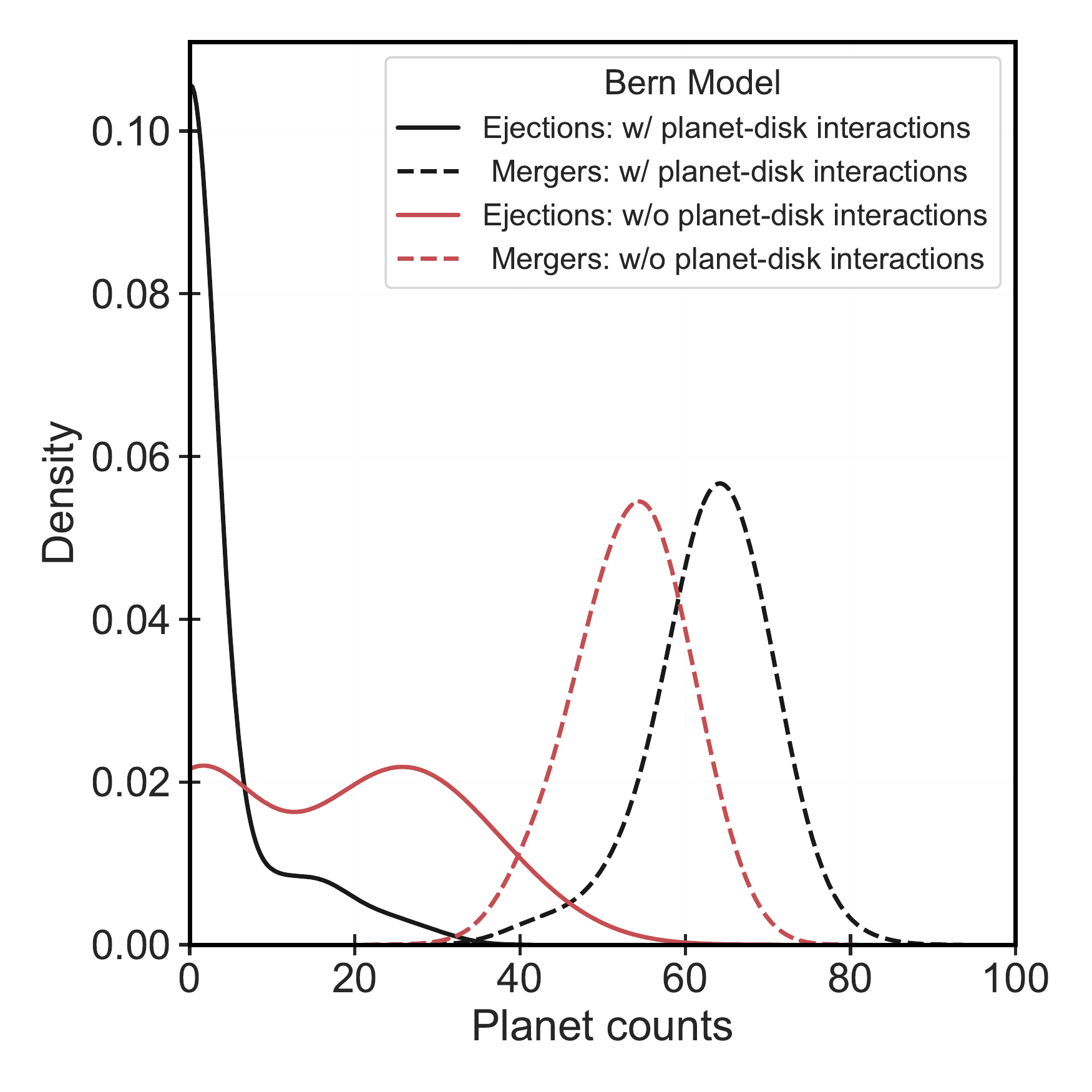}
		\caption{Effect of planet--disk interactions on architecture. Left: Scatter plot shows the number of planet--planet mergers and planetary ejections that occurred in systems as a function of disk lifetime. The solid lines show the average counts for each disk lifetime bin. Right: Distribution of the total number of mergers (dashed) and ejections (solid) for the entire synthetic population. The black line depicts the nominal synthetic population, and the red line depicts a different synthetic population in which the disk-)planet interactions were artificially switched off.}
		\label{fig:diskplanet}
	\end{figure*}
	
	Protoplanetary disks interact with planets via several mechanisms. Planets may experience orbital migration via gravitation interactions with the disk.  Low-mass planets undergo type I migration, which in the Bern model is implemented following the approaches of \cite{Coleman2014, Paardekooper2011}. Massive planets may open a gap in the disk and undergo type II migration \citep{Dittkrist2014}. The disk also dampens the eccentricity and inclination of planets, which is coherently applied within the \textit{N}-body integrator. Readers interested in the details of the implementation are referred to \cite{Emsenhuber2021A, Emsenhuber2021B}. 
	
	Figure \ref{fig:diskplanet} (left) shows the count of mergers and ejections for each planetary system in our synthetic population as a function of the  lifetime of its protoplanetary disk. For an easier visualisation of any underlying trend, we also show the average merger and ejection counts for each disk lifetime bin. The number of planet--planet mergers shows a clear correlation with disk lifetime. Disks that live longer usually give rise to planetary systems that undergo more mergers than short-lived disks. We refer to this correlation as `migration assisted mergers'. One possible explanation for this correlation could be that disks allow planets to migrate depending on their mass \footnote{There could be other scenarios which contribute to the \change{`migration assisted mergers'} correlation. For example, migration may allow planets to become more massive by accreting more material due to increased access to planetesimals \citep{Alibert2005}. Massive planets may interact more amongst themselves, leading to more mergers.}. Two adjacent planets that are not migrating at the same rate, perhaps owing to their different masses, can come close enough for a merger to occur. The number of ejections does not show any clear trend with disk lifetime. Disks dampen a planet's eccentricity and inclination. As ejection requires extremely violent interactions (marked by high eccentricities and inclinations), disks may essentially inhibit planetary ejections.
	
	To test these ideas, we simulated another population of 1000 planetary systems. In this population (NG140), planet--disk interactions (gas-driven migrations, and eccentricity and inclination damping) are artificially switched off. For all such systems, we count the number of mergers and ejections and compare them with our nominal population. Figure \ref{fig:diskplanet} (right) shows the distribution of the number of planet--planet mergers and planetary ejections in the two populations. 
	
	As expected, the number of planet--planet mergers decreases (distribution shifts to the left) when planet--disk interactions are switched off. This confirms the \change{migration-assisted mergers} correlation presented above. The distribution of ejections, on the other hand, increases significantly when planet--disk interactions are switched off. When the damping of the planetary eccentricity and inclination by the disk is switched off, the gravitational interactions between planets increases, such that many planets are ejected.
	
	We make two observations from the results presented so far.
	First, counts of mergers and ejections seem to be a good proxy for the prevalence of dynamical interactions, as they capture some of the well-established dynamical effects concerning planet--disk interactions. Second, we observe that disks affect system architecture in a multitude of ways. While disk mass shows a direct relation to final architecture, disks also affect system architecture indirectly by influencing the dynamical interactions that occur therein. Long-living disks give rise to more mergers and inhibit planetary ejections. Conversely, systems emerging from short-lived disks experience fewer mergers.

	\subsection{Planet--planet interactions}
	
	\def\figwidtha{4.5cm}
	\begin{figure*}[]
		\centering
		\includegraphics[width=\figwidtha]{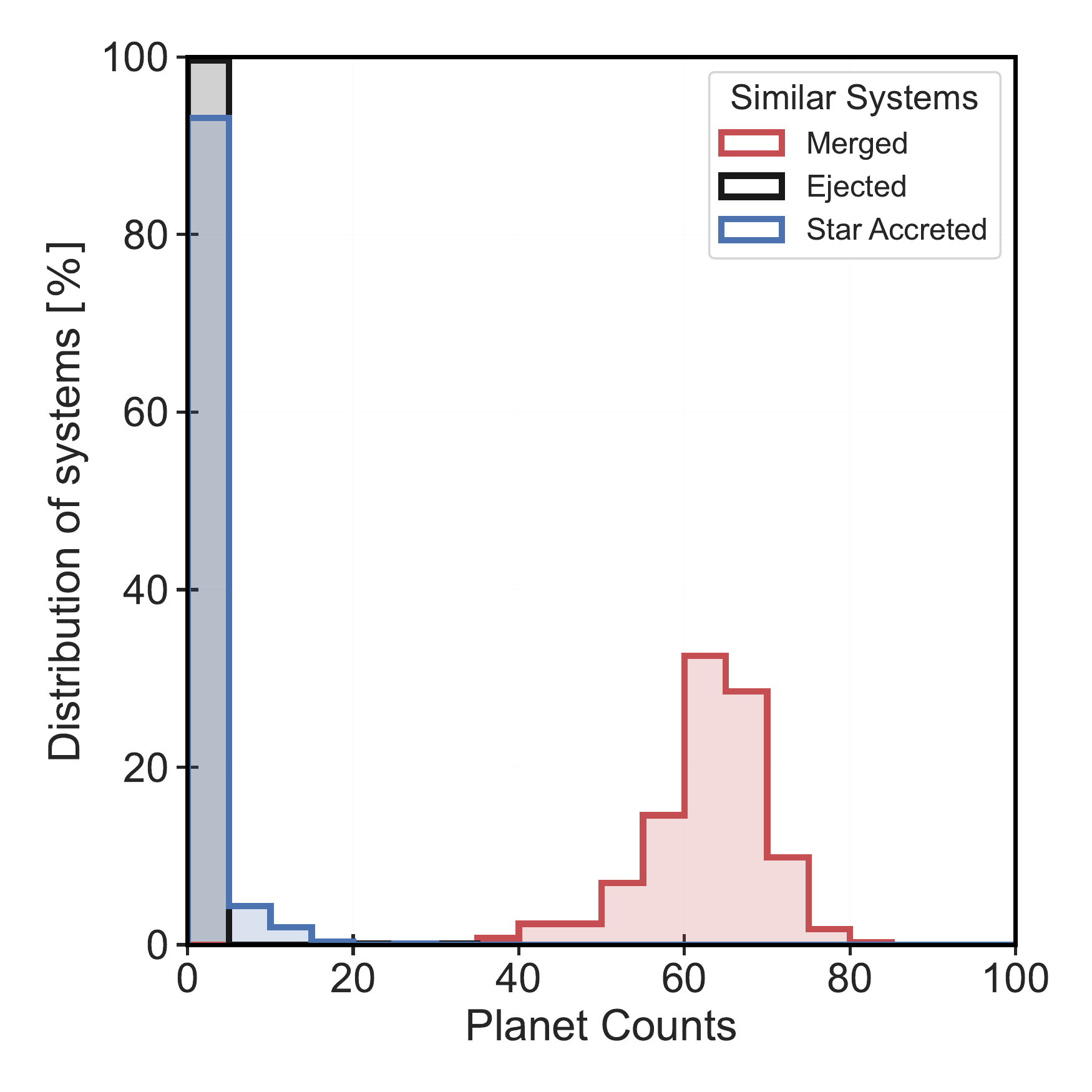}           
		\includegraphics[width=\figwidtha]{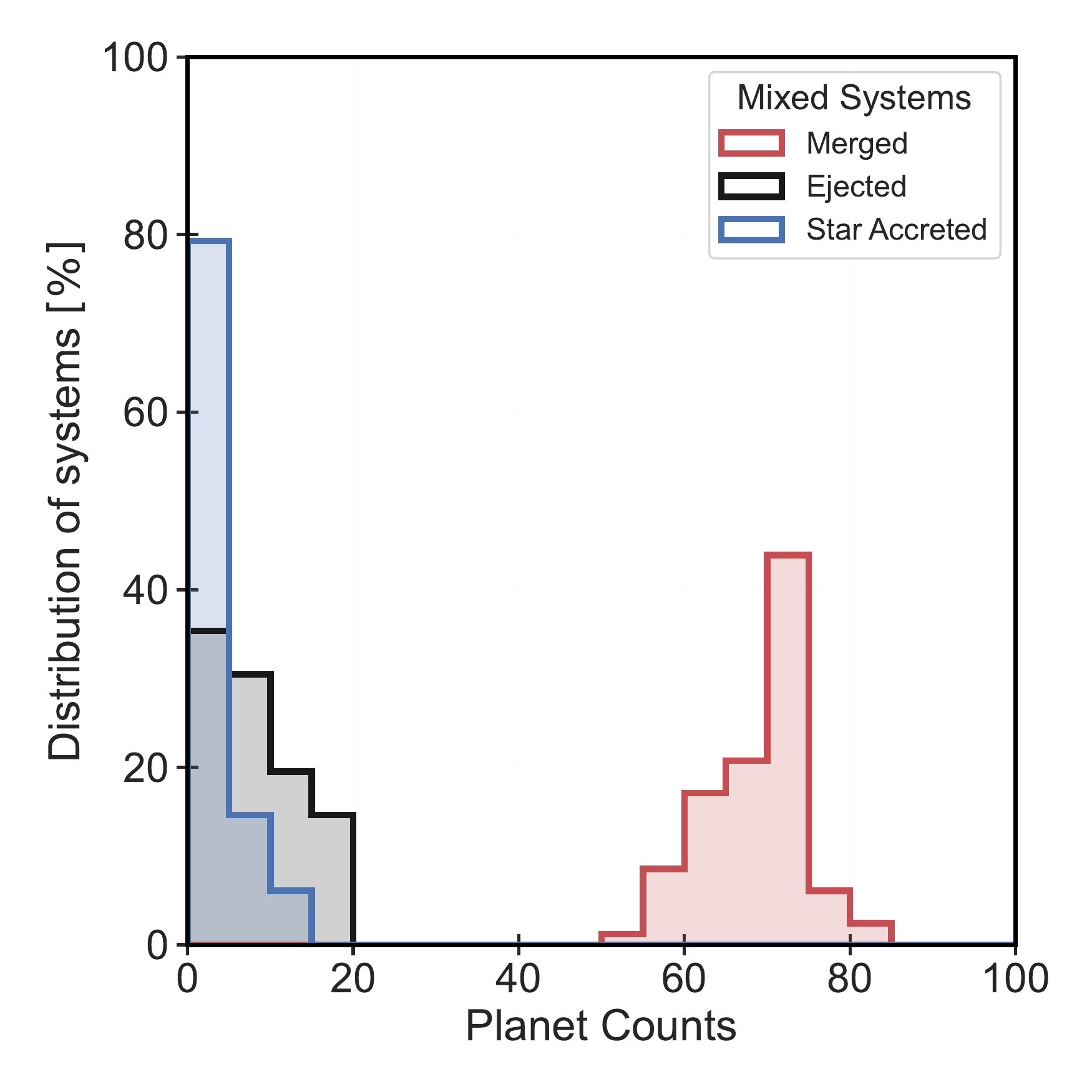}
		\includegraphics[width=\figwidtha]{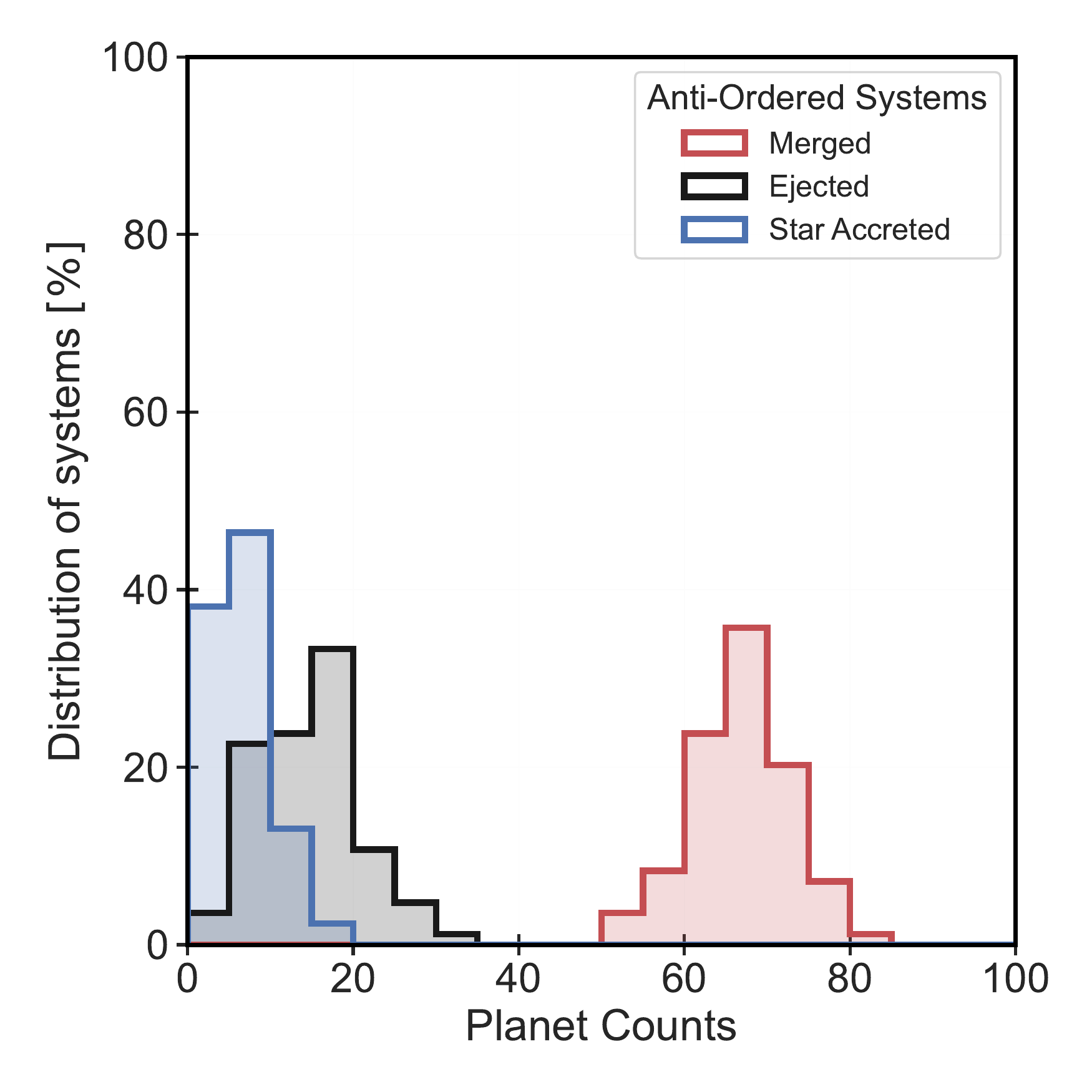}
		\includegraphics[width=\figwidtha]{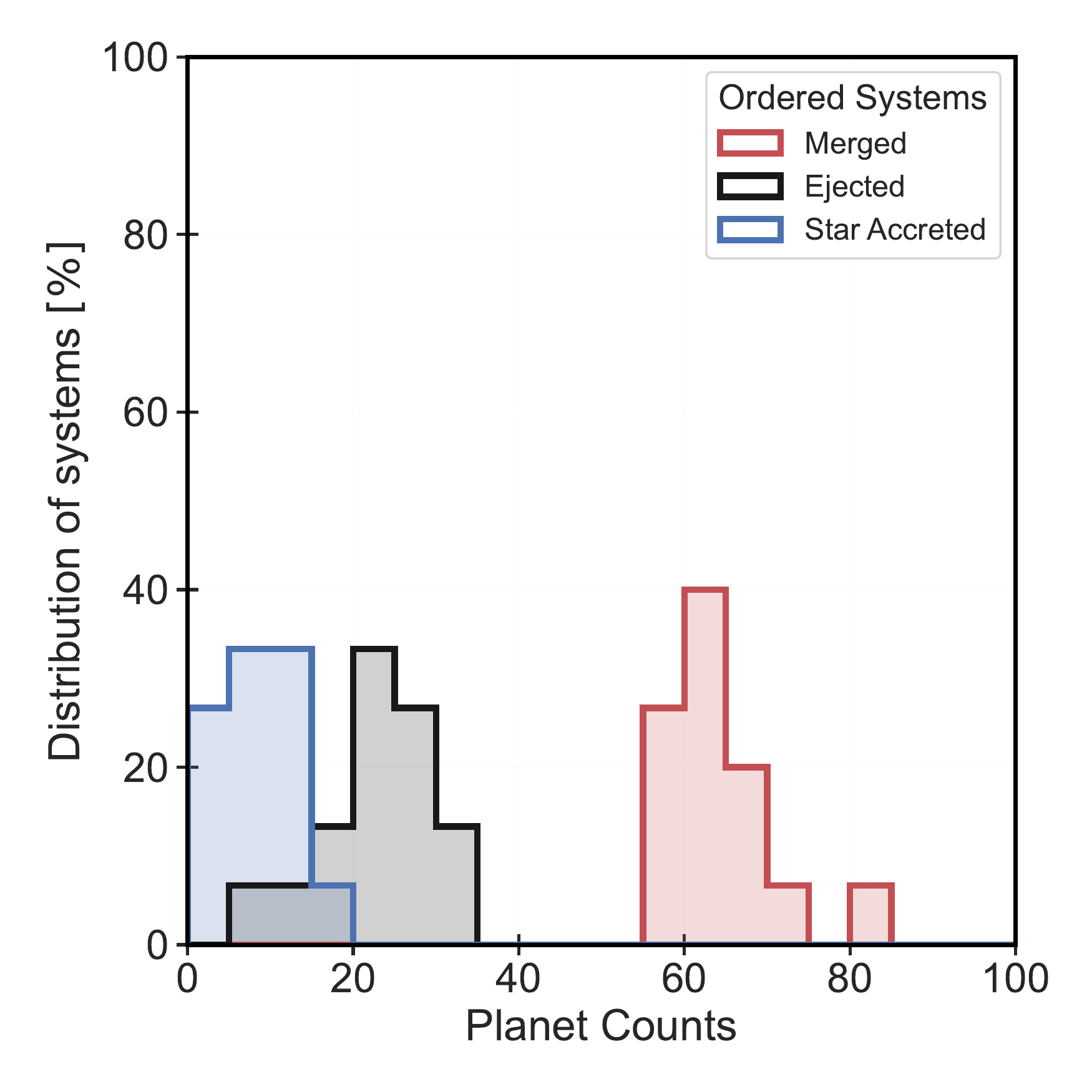}
		\caption{Effect of planet--planet interactions on system architecture. For each architecture class, the panels show a histogram of the counts of planet--planet mergers, ejections, and stellar accretion occurring in the synthetic population. The y-axis in all panels is scaled to reflect the percentage of systems in each of the four architecture classes. For example, $100\%$ of all \similar/ systems lost less than five planets via planetary ejection.}
	\label{fig:planetplanet}
\end{figure*}

Above, we show that planet--disk interactions in the Bern model may influence the dynamical interactions occurring in a system. Now, in this section, we are interested in understanding how these violent events shape the final architecture of a system. 

Planets interact with each other gravitationally. These multi-body interactions are tracked via a \textit{N}-body integrator in the Bern model. The end result of some of the more violent interactions is that planets are lost via one of several channels:  planet--planet mergers\footnote{In our model, when the distance between two planets becomes smaller than the sum of their radii, a planet--planet collision is said to occur. We treat such merger events in a simplified manner: the cores of the target--impactor pair are merged, the lesser massive body loses its envelope, and the impact energy is added to the merged new body following \cite{2012A&A...538A..90B}, which determines what part of the gaseous envelope is ejected.}, planetary ejections, accretion by the host star, and so on. These channels allow a planetary system to fundamentally alter itself and its architecture. 

Figure \ref{fig:planetplanet} shows, for each architecture class, the distribution of planet--planet mergers and the number of planets lost via ejections and stellar accretion. At first glance, losing planets to the host star may not seem appropriate for planet--planet interactions. However, many of these planets meet their fate, in the Bern model, when they are pushed inwards after being captured in mean-motion resonances with other planets\footnote{The model also includes inward migration of planets as a result of the stellar tides.}. Therefore, this channel of losing planets is included here. We caution the reader that the absolute number of planets lost via any channel is model-dependent. The quantity of interest here is the relative difference between the different architecture classes.

Figure \ref{fig:planetplanet} suggests that the \similar/ architecture class is almost completely shaped by planet--planet mergers. Most \similar/ systems in our simulations have between 40 and 80 mergers taking place within them, and the median number of mergers is 63. Violent dynamical interactions that lead to the ejection of planets seems to be very rare in this architecture type, as $100\%$ of all \similar/ systems lose less than five planets via planetary ejection (median ejections is 0). Likewise, \similar/ systems seem to not rely on the stellar accretion channel for losing planets (median stellar accretions is 0).

Systems with \mixed/ architecture \change{also} undergo many planet--planet mergers. The number of mergers in \mixed/ systems ranges from 50 to 85, and the median of mergers is 70. In a clear contrast from \similar/ architectures, the ejection and stellar accretion channels play an important role for \mixed/ systems. The median number of planets lost via ejections is 7, and via stellar accretions is 2. 

\antiordered/ systems utilise all three dynamical channels. The distribution of mergers in \antiordered/ systems is roughly similar to that of \mixed/ systems. The range is between 50 and 85 and the median number of mergers is 67. However, \antiordered/ systems tend to lose more planets via the ejection channel. The number of planets lost via dynamical ejection ranges from 0 to 35 with a median value of 14.5. Compared to \mixed/ systems, \antiordered/ systems also tend to lose more planets via stellar accretion (median is 6).

Amongst the four architecture classes, \ordered/ systems seem to undergo the greatest number of dynamical interactions.  The distribution of planet--planet mergers in \ordered/ systems shows a tail-like feature. The number of mergers ranges from 55 to 85, with 62 being the median. All \ordered/ systems eject at least five planets. The number of ejections has a range from 5 to 35, and the median is 23. The distribution of planets lost via the stellar accretion channel shows a shift to the right. The number of planets accreted by the star ranges from 0 to 20 with 8 being the median. 

A comprehensive picture of the role of dynamical history in shaping the final architecture emerges from the four panels in Fig. \ref{fig:planetplanet}. Similar systems tend to rely only on the merger channel for shaping their system architecture. As planetary systems in all four architecture classes undergo a considerable number of mergers, this channel may not suffice to explain or distinguish the emergence of the four architecture classes. This is in line with what was found before, namely that the emergence of the \similar/ class is mostly governed by the initial conditions. 

While initial conditions seem to decide whether a system becomes \similar/ or one of the other three architectures, there appears to be a trend in the role of dynamical interactions in shaping \mixed/, \antiordered/, and \ordered/ architectures. \change{The distributions of the ejection and accretion channels distinguish these three architectures. These distributions show a shift to the right, indicating that more planets are being lost via these two channels as we move from \mixed/ to \antiordered/ and to \ordered/ architectures.} Thus, we conclude that if initial conditions do not allow a system to become \similar/, its fate is decided by its dynamical history, among other effects. If the strength of the dynamical interactions increases in a system, the architecture of the system changes from \mixed/ to \antiordered/ or to \ordered/. 

All systems in the Bern model start with 100 protoplanetary embryos. Above, we show that systems of different architecture show varying propensity to lose planets via the different dynamical channels. This suggests that we should also see an effect of the dynamical history of the four architecture classes in their multiplicity distribution. We observed
this effect in Fig. 6 of \paperone/. We do not have a way to determine the initial
number of embryos of the planetary systems we observe today. Our approach may therefore not be directly applicable to observed planetary systems. We remind the reader that while the quantitative aspects we present in this section are probably model dependent, the qualitative nature of these results is of paramount importance.

\section{The Aryabhata formation scenario}
\label{subsec:mohicans}

\change{In this section, we propose a planet-formation scenario to explain a feature observed by \paperone/ (Sect. 5.4). We found that  many synthetic planetary
	systems have a peculiar water-mass-fraction architecture namely that all planets hosted in these systems are water-rich worlds. We explain this peculiar feature with the `Aryabhata formation scenario'.}

\change{
	The first exoplanets to be discovered were hot Jupiters ---giant planets orbiting their host stars at very short periods \citep{Mayor1995}. Orbital migration was suggested as a possible mechanism to explain these short periods \citep{Lin1996, Lin1997}. Theoretical studies indicate that orbital migration and planet--star tidal interactions should make many close-in planets unstable. In the 1990s, Doug Lin described `the last of the Mohicans' scenario \citep{Garaud2011}. In this scenario, the protoplanetary disk gives rise to planets, many of which are doomed to fall onto the star. The surviving observable planets are those that were able to escape annihilation.
}

\change{
	For some simulated systems, we noticed a modified version of this scenario. Protoplanetary disks seem to give rise to planets at different epochs. In the first epoch, several intermediate-mass planets ($1-100 \mearth$) are formed within the first 1Myr. Most of these `first generation' planets are subsequently lost mainly via giant impacts (and a few are lost via orbital or tidal migration leading to stellar accretion). This purging phase is catastrophic to all planets that started within the ice line. Over the next few million years, a second epoch sees the advent of a `second generation' of planets. Most of these second-generation planets are born outside the ice line, and are able to migrate inwards during the disk lifetime. After disk dissipation, migration comes to a halt and many of these planets survive long-term \textit{N}-body evolution in our simulations. We call this the Aryabhata formation scenario. The key difference between the two scenarios is that in the Aryabhata formation scenario (a) planets (surviving and lost) are born in different epochs, and (b) most first-generation planets are lost via giant impacts. 
}

\change{
	We quantify this scenario with the Aryabhata's number, $\mu$, which is the ratio of the surviving planets that started inside the ice line to the total number of surviving planets: 
}
\begin{equation}
	\text{Aryabhata's number:}\ \mu = \frac{\nplanet (a^\text{start}_\text{embryo} \leq a_\text{ice})}{\nplanet}
	.\end{equation}
\change{
	At the start of our calculations, all systems have an Aryabhata's number $\approx 0.5\pm0.1$. Figure 12 of \paperone/  (middle) shows the ice mass fraction architecture of simulated planetary systems. The colour of each point shows the Aryabhata's number. 
}

\change{
	Most planetary systems with $\cs(f_{ice}) \approx \cv(f_{ice}) \approx 0$ have $\mu$ close to zero. This suggests that most (or all) of the surviving planets in such systems started outside the ice line. The formation path of these systems falls into the Aryabhata formation scenario. These classes of systems can be identified by two characteristics: (i) the core water-mass fraction for different planets in these systems is similar, and (ii) the core water-mass fraction for most planets is high (owing to their  origin outside the ice line) making them water-rich planets. Approximately, one-fifth of the simulated systems fall into this scenario. Among these, about half are of \similar/ class, one-third are \antiordered/, and the remaining systems have either a \mixed/ or \ordered/ mass architecture. 
}

\change{
	There exists an almost linear relationship between $\cv(f_{ice})$ and $\mu$. Using scipy's linear regression module, we obtain a slope of 1.8 and intercept of 0.18 between these two quantities. The correlation coefficient is $R = 0.95$, indicating a strong correlation between the Aryabhata's number and the \cvtext/ of core water mass fraction. This suggests a possibility to identify observed exoplanetary systems that may have originated via the Aryabhata formation scenario. By determining the $\cv(f_{ice})$ of a system, the Aryabhata's number can be estimated. Systems with low $\mu$ values probably arose from this scenario. 
}

\change{
	For systems that fall into the default scenario (positive $\cs(f_{ice})$, implying an increasing core water mass fraction inside-out), the Aryabhata's number is $\mu >0$. We note that most systems with $\mu \gtrapprox 0.6$ show similarity in their mass architecture. 
}

\change{
	Overall, the intra-system core water-mass-fraction architecture of most planetary systems seems to take one of two forms. (i) Those characterised by $\cs(f_{ice}) \approx \cv(f_{ice}) \approx 0$ and $\mu = 0$. These systems are composed of water-rich planets wherein the core water mass fraction is similar across the different planets. All surviving planets in these systems started outside the ice line. The Aryabhata formation scenario explains these systems. (ii) Those with $\cs(f_{ice}) > 0$ and $\mu > 0$. These systems represent the `default' or common outcome of our simulations. The planetary core water-mass fraction in these systems increases from one planet to another with increasing distance from the host star. Some of the surviving planets started from inside the ice line. At the extreme end, systems in which $60\%$ or more surviving planets started inside the ice line tend to have a similar mass architecture. 
}

\section{Summary, conclusions, and future work}
\label{sec:conclusions} 

\paperone/ of this series introduced a novel, model-independent framework for characterising the architecture of planetary systems at the system level. Planetary-system architectures can be separated into four classes: \similar/, \mixed/, \antiordered/, and \ordered/. This classification is achieved via two quantities: the \cstext/ and the \cvtext/. The mathematical $\cs$ versus $\cv$ architecture space was found to have forbidden regions -- regions in which no planetary system can exist. In \paperone/, the mass architecture classes of observed and synthetic systems were characterised. The mass architecture of synthetic systems was compared with their radii architecture, bulk-density architecture, core-mass architecture, spacing architecture, and water-mass-fraction architecture. \change{As in \paperone/, we identify a system's architecture with its mass architecture.}

In this paper, we explore the core-accretion-based formation pathways ---around a solar-like star--- of the four classes of planetary system architecture. We tried to disentangle the role of {nature} (initial conditions of planet formation) from that of {nurture} (physical processes occurring during planet formation). Our findings can be summarised as follows:

\begin{enumerate}
	
	\item \textbf{System-level analysis:} \change{Our findings show that a system-level analysis of planetary system architecture via our architecture framework (\paperone/) provides an abundance of information. We show that planetary formation and evolution process leave their imprint on the entire system architecture. }
	
	\item \textbf{Solid disk mass:} The initial amount of solids in the protoplanetary disk in our models plays an important role in deciding the architectural fate of a planetary system. Disks with a solid mass (initial content of planetesimals) of $\lesssim 1 \mjupiter$ almost always give rise to systems with \similar/ architecture. Mixed architectures arise most often from disks with solid masses $\approx 1 \mjupiter$. Disks with solid mass $\gtrsim 1 \mjupiter$ favour the production of \antiordered/ and \ordered/ architectures. 
	
	\item \textbf{Gas disk mass and metallicity:} Initial gas disk mass and stellar metallicity influences the final architecture of a planetary system by controlling the initial mass of solids in the disk. Metallicity, in our models, is simply related to the dust-to-gas ratio, which allows us to convert a fraction of the initial gas disk mass into initial dust mass (eq. \ref{eq:fpg}). Applying the architecture framework on the synthetic systems from the Bern model allows us to predict the existence of a metallicity--architecture correlation. The observed correlation between metallicity and final architecture is in qualitative agreement with the Bern model. 
	
	\item \textbf{Metallicity--architecture correlation:} The architecture of a planetary system correlates with the metallicity of the host star. Most systems hosted by a low-metallicity star ($Fe/H < - 0.2$) are of \similar/ architecture. As the metallicity of the star increases, \mixed/, \ordered/, and \antiordered/ architectures become increasingly common.
	
	\item \textbf{Disk lifetime:} The occurrence of systems of a \similar/ architecture around short-lived disks is high, and their frequency reduces around long-lived disks. The frequency of \antiordered/ architecture increases as disk lifetime increases. These correlations are mediated in at least two ways. First, disks interact with planets, where orbital migration and eccentricity and inclination damping occur. Due to the `migration assisted merger' correlation, long-lasting disks allow planetary systems to have,  in general, more planet--planet mergers and inhibit planetary ejections. These dynamical events shape a system's final architecture. In addition, in our model, disk lifetimes are correlated with disk masses, which also strongly influences the system architecture.
	
	\item \textbf{Dynamical interactions:} Planetary systems can significantly alter their architecture via (at least) three dynamical channels: planet--planet mergers, planetary ejections, and accretion by the host star. All architecture classes in our formation model were found to undergo numerous merger events. Similar systems rely entirely on mergers to shape their final architecture. As the strength of the dynamical interactions experienced by a system (quantified by the number of ejections and/or accretions) increases, the architecture of a system shifts from \mixed/ to \antiordered/ to \ordered/.

	\item \change{\textbf{The Aryabhata formation scenario}: Systems following this formation scenario have the following formation pathway. First-generation planets (formed within 1 Myr) are lost mostly via giant impacts. Second-generation planets started outside the ice line and migrated inwards. The surviving planets are from the second generation and shape the architecture of the system. This scenario explains about $20\%$ of simulated systems in which the core water-mass-fraction architecture is different from the default scenario. Systems following this formation scenario (i) host only those planets that have a high core water-mass fraction and (ii) host only those planets that started outside the ice line. We introduce the Aryabhata's number to identify those systems that follow this formation scenario and find that $80\%$ of all \antiordered/ simulated systems are formed via the Aryabhata formation scenario.}
	
	\item \textbf{Nature versus nurture:} Overall, our model suggests that initial conditions ---or {`nature'---} dictate whether a system will have a \similar/ architecture or one of the other three architecture classes, namely \mixed/, \antiordered/,
	or \ordered/  (via initial disk mass). If {nature} does not allow a system to have a \similar/ mass architecture, then the final architecture is controlled by {`nurture',} or dynamical interactions, among other possible effects. As the dynamical interactions increase, the final architecture tends to become \mixed/, \antiordered/, and then \ordered/.
\end{enumerate}

We would like to offer readers some warning when interpreting our results. Although the architecture framework (from \paperone/) is model-independent, the present results hinge critically on the underlying planet formation model -- the Bern model. There are several assumptions, simplifications, and choices to be made when simulating synthetic planetary systems using the Bern model. For example, the treatment of planet--planet merging collisions is relatively simple \citep{2022MNRAS.509.1413A}. We also assume simplified planet-formation conditions; that is, our star--disk--planet system is isolated enough so that we may ignore the influence of the stellar neighbourhood, stellar flybys, and so on \citep{2012MNRAS.419.3115B, 2018MNRAS.475.5618B}. The main strength of this study does not lie in providing an explanation of the formation pathway of any particular system. Instead, our main result is the observation that when groups of planetary systems are identified (architecture classes), general trends in formation pathways emerge. This allowed us to map the roles of {nature} and {nurture} in shaping the final architecture of a planetary system. 

The results of this study can be strengthened or challenged in several observational and theoretical ways. We list some possibilities for future studies emerging from this work:

\begin{enumerate}
	\item \textbf{Linking disk mass distribution and architecture occurrence rates:}
	Our model suggests that there should be a direct relationship between the mass of the solid disk and the final architecture of a system. While initial disk masses and the final architecture of the same system will forever remain unobservable, this relation can be tested statistically. The distribution of initial disk masses and the distribution of final system architecture can be linked by formation models. We speculate that in future, when these two distributions become available, formation models can be used to predict one or the other. In fact, this problem can also be turned around; we can identify the right family of models as those that correctly link the observed distributions of protoplanetary disk masses and architecture occurrence rates. We believe such tests are crucial for the development and eventual emergence of a standard model for exoplanetary astrophysics. 
	
	\item \textbf{Metallicity--architecture correlation:}
	Our work suggests that the current architecture of a planetary system should be related to the metallicity of its  host star. As both of these are observable, testing this metallicity--architecture correlation should be feasible. Here, we used a catalogue of 41 observed multi-planet systems (from \paperone/) to test this correlation. We find a qualitative agreement between theory and observations. However, our observational catalogue suffers from incompleteness and low-number statistics, which prevents us from making any further assertions. More observational data are required to confirm or reject the proposed metallicity-architecture correlation. It would also be interesting to estimate the current architecture occurrence rate based on the known metallicity distributions. 
	
	\item \textbf{Confirming formation pathways:} 
	Confirming the formation pathways discovered in the present study with observations is challenging. However, the strength of our results will increase if different planet-formation models are studied through the architecture framework. Hence, one possible line of future work involves repeating the present study using different planet-formation models. 
	
	\item \textbf{Extending the architecture framework:}
	So far, we have calibrated our classification scheme for the mass architectures only. Calibrating the architecture classification framework on other quantities maybe useful. Especially for planetary radii, which are observable via transit surveys, the use of machine learning methods may be necessary. 
	
	\item \textbf{Temporal evolution of system architecture:}
	In the nominal Bern model population studied in this paper, protoplanetary embryos of  100 lunar masses  are initialised in the protoplanetary disk at the start. This necessarily implies that all planetary systems start as \similar/ type systems. It would be interesting to inquire whether this is generally true in nature as well.  If this is the case, this implies that the `default' architecture of all planetary systems is \similar/ and the physical processes playing out in the system evolve this architecture into other possibilities. Investigating this may lead to deep insights into the structure of planetary system architecture. In addition, such studies would be necessary to interpret the observed architecture occurrences, as observed planetary systems are seldom of the same age. 
	
	\item \textbf{External perturbations:} Stellar flybys or multi-planetary systems around binaries provide excellent theoretical and observational laboratories with which to study the influence of external perturbations on the architecture of planetary systems. This problem, when turned around, is also useful in deducing the perturbed or dynamical (or lack of) history of observed planetary systems.
	
\end{enumerate}

This paper presents new insights obtained by analysing planetary systems at the system-level. We showed that several patterns emerged in the formation pathways of the four architecture classes. These patterns linked the initial conditions of planet formation with the final architecture of a system -- bridging the vast temporal gap of several billions of years between the birth of planets to their final assembly.

\begin{acknowledgements}
	
	This work has been carried out within the frame of the National Centre for Competence in Research PlanetS supported by the Swiss National Science Foundation.  We acknowledge the support of the Swiss National Fund under grant 200020\_172746 and 200021\_204847 ``PlanetsInTime''. LM acknowledges the generous hospitality of the "Planet Formation" workshop by the Munich Institute for Astro-, Particle and BioPhysics (MIAPbP) which is funded by the Deutsche Forschungsgemeinschaft (DFG, German Research Foundation) under Germany's Excellence Strategy – EXC-2094 – 390783311.
	\\
	\textit{Data:} The synthetic planetary populations (NGPPS) used in this work are available online at \url{http://dace.unige.ch}. \textit{Software:} Python \citep{python3}, NumPy \citep{numpy}, Seaborn \citep{seaborn}, Pandas \citep{pandas}, Matplotlib \citep{matplotlib}.
	
\end{acknowledgements}

\bibliographystyle{aa}
\bibliography{exoplanets}

\end{document}